\documentclass[a4paper, amsfonts, amssymb, amsmath, reprint, showkeys, nofootinbib, twoside, superscriptaddress]{revtex4-2}
\usepackage[english]{babel}
\usepackage[utf8]{inputenc}
\usepackage[colorinlistoftodos, color=green!40, prependcaption]{todonotes}
\usepackage{amsthm}
\usepackage{amsmath}
\usepackage{mathtools}
\usepackage{physics}
\usepackage{xcolor}
\usepackage{graphicx}
\usepackage[left=18mm,right=18mm,top=35mm,columnsep=15pt]{geometry} 
\usepackage{adjustbox}
\usepackage{placeins}
\usepackage[T1]{fontenc}
\usepackage{lipsum}
\usepackage{csquotes}
\usepackage{siunitx}
\usepackage{caption}
\usepackage{subcaption}
\usepackage[pdftex, pdftitle={Article}, pdfauthor={Author}, colorlinks=true]{hyperref}
\begin{document}
\title{ 
Rotational stability in nanorotor and spin contrast in one-loop interferometry in the Stern-Gerlach setup
}

\author{Ryan Rizaldy}
    \affiliation{Van Swinderen Institute for Particle Physics and Gravity, University of Groningen, 9747AG Groningen, the Netherlands}

\author{Tian Zhou}
    \affiliation{Van Swinderen Institute for Particle Physics and Gravity, University of Groningen, 9747AG Groningen, the Netherlands}

\author{Sougato Bose}
    \affiliation{Department of Physics and Astronomy, University College London, London WC1E 6BT, United Kingdom}

\author{Anupam Mazumdar }
    \affiliation{Van Swinderen Institute for Particle Physics and Gravity, University of Groningen, 9747AG Groningen, the Netherlands}


\begin{abstract}
The rotation of a nanoparticle in a quantum system has many applications, from theory to experiments. This paper will treat nanoparticle rotational dynamics for spin-embedded nanorotors. We will model it as a rigid body that properly treats the rotation in the co-frame of the nanorotor in the presence of external fields. Besides rotation, we will further investigate how to create large spatial superpositions in the inhomogeneous external magnetic field, such as in the case of the Stern-Gerlach apparatus. The spin-embedded nanorotors play a crucial role in creating matter-wave interferometers through their spin and external magnetic field interaction Hamiltonian.  We aim to provide a holistic interpretation of the dynamics of three Euler angles, their quantum evolution, and the nanorotor's spatial motion in a Stern-Gerlach-type setup where we will consider one-full-loop interferometry. We will then study how the quantum evolution of all the Euler angles leads to a spin coherence loss upon interference and what manifests the Einstein-de Haas effect in an external magnetic field. In particular, we show that by imparting rotation along the direction of the magnetic field, we can stabilise the nanorotor's libration mode. We will also extend our analysis to a case where the initial state of the libration mode is thermal and discuss the contrast loss due to interference of the nanorotor upon one-loop completion.

\end{abstract}

\maketitle


\section{INTRODUCTION}

There are many applications for creating spatial superposition with nanoparticles, especially embedded with an electronic spin defect center, ranging from quantum metrology to quantum sensors. One such particular spin defect will be nitrogen vacancy (NV)-centered nanodiamond~\cite{Doherty_2013}. One particular aspect of creating superposition lies in fundamental physics by applying the Stern-Gerlach force to create spatial matter-wave interferometry~\cite{margalit2018realization,SGI_experiment,amit2019t}. We can apply matter-wave interferometer to test the foundations of quantum mechanics~\cite{bassireview}, create quantum sensors~\cite{Wu:2022rdv,Wu:2024bzd,Toros:2020dbf}, detect dark matter/axion particles~\cite{Barker:2022mdz}, gravitational waves~\cite{Marshman:2018upe}, equivalence principle~\cite{Bose:2022czr}, quantum analogue of the light bending experiment~\cite{Biswas:2022qto}, test modified theories of gravity~\cite{elahi2023probing,Vinckers:2023grv,Chakraborty:2023kel}, and test the quantum nature of spacetime (the quantum gravity-induced entanglement of masses (QGEM) experiment)~\cite{Bose:2017nin,ICTS}, see also~\cite{Marletto:2017kzi}. The latter also pushed the low energy frontier to test the relativistic modifications to quantum electrodynamics~\cite{Toros:2024ozf} and post-Newtonian corrections to quantum gravity~\cite{Toros:2024ozu}. 

Significantly, the QGEM protocol requires two adjacent massive superpositions with the help of spin embedded in nanoparticles~\cite{Bose:2017nin,ICTS}, see also theoretical aspects of QGEM~\cite{marshman2020locality,Bose:2022uxe} and \cite{Carney_2019,Carney:2021vvt,Danielson:2021egj,christodoulou2023locally}. The entanglement due to the quantum nature of gravity can be witnessed by reading the spin of nanoparticle~\cite{Bose:2017nin,Schut:2021svd,Tilly:2021qef}. Furthermore, we must mitigate photon-mediated entanglement based on Casimir and dipole interactions~\cite{vandeKamp:2020rqh,marshman2024entanglement}~\footnote{The original idea was indeed closely related to the story of NV-centered nanodiamond, but the current analysis holds true for any electronic spin embedded in a nanoparticle, with diamagnetic properties similar to that of a nanodiamond. Hence, we will be generic and use the word nanorotor, but for numerical purposes, we will use the properties of a nanodiamond.}.

Nevertheless, creating a quantum spatial superposition of distinct localized states of neutral objects remains a significant challenge theoretically and experimentally. QGEM protocol requires $m \sim 10^{-14} - 10^{-15}$ kg over spatial separations of $\Delta x \sim {\cal O}(10-50) {\rm \mu m}$~\cite{Bose:2017nin,Schut:2023hsy,Schut:2023eux}, far beyond the scales achieved to date (e.g., macromolecules $m \sim  10^{-22}$ kg over $\Delta x \sim 0.25 {\rm \mu m}$, or atoms $m\sim 10^{-25}$ kg over $\Delta x\sim 0.5 \ \mu\text{m}$ m~\cite{arndt,overstreet2022observation,asenbaum2017phase}. Furthermore, there are additional effects, such as the rotational effects of nanorotors, which in the presence of the external magnetic field leads to external torque, libration of the NV spin, and the famous Einstein-de Haas effect,~see~\cite{stickler2021quantum} and \cite{Japha:2022phw, japha2021unified}, the superposition size also depends on the nature of the spin states of the two arms, hence on the spin-libration degrees of freedom. The rotational effects are known to be important in nanocrystals; see these very important contributions ~\cite{chen2019nonadiabatic,Stickler18_GM,Rusconi:2022jhm,stickler2021quantum,ma2021torque,jin2024quantum,Kuhn2017}. It was also pointed out that rotational effects also modify the size of the superposition and the spin contrast~\cite{Japha:2022phw}, known as the Humpty-Dumpty problem \cite{Englert,Schwinger,Scully}. The problem states that when the two trajectories meet, their classical positions, momenta, and any other rotational degrees of the rigid body (such as the three Euler angles related to the NV spin) should match along with their quantum wave packets and must significantly overlap so that significant spin contrast can be achieved in an experiment.
Stern-Gerlach-based experiments are based on the spin readout, which is extremely promising in NV-centred nanodiamonds~\cite{WoodPRA22_GM,WoodPRB22_GM}. This is the only reason the experiments rely on the final spin measurement. After completion of interference, one should combine the spatial superposition of the nanodiamonds and consider the coherence (or contrast) loss of the NV spin~\cite{Japha:2022phw}.

There are many schemes that create a spatial superposition in the Stern-Gerlach setup, see~\cite{WanPRA16_GM,Scala13_GM,Pedernales:2020nmf,Marshman:2021wyk,Zhou:2022epb,Zhou:2022frl,Zhou:2022jug,Zhou:2024voj}. However, all of these schemes suffer from the Humpty-Dumpty problem. The idea is to minimize Humpty-Dumpty to ensure that visibility loss is minimal. The cause of this problem could be the slightest mismatch in the classical trajectories at the end of the experiment when the spin readout is made. In general, we can imagine three sources of spin contrast loss: (1) the decoherence effect of the interaction between the nanorotor and the environment, such as collisions by air molecules, electromagnetic noise, and spin-spin interaction between the spin defect and another spin in the nanoparticle or the environment~\cite{Bose:2017nin,vandeKamp:2020rqh,Rijavec:2020qxd,Schut:2021svd,Schut:2023hsy,Schut:2023eux,Fragolino:2023agd,Schut:2023tce, Schut:2024lgp}. (2) Noise in the current, such as current fluctuation, which controls the magnetic field and the gradient~\cite{Marshman:2018upe,
Zhou:2022epb,Zhou:2022frl,Zhou:2022jug,Zhou:2024voj}. (3) the fluctuations in the Euler angles in the rigid body, which experiences an inhomogeneous magnetic field in the Stern-Gerlach setup. A small manifest of this third problem was studied in the context of fluctuations in the libration mode in these very important papers~\cite{Japha:2022phw, japha2022role}. The libration mode represents the misalignment angle between the nanodiamond's NV axis and the external magnetic field, although the analysis is very general and can be applicable to any spin system in a rigid body. The latter is inevitable in a nanodiamond since the NV spin state $s=1$ yields non-vanishing zero-field splitting of the NV spin, which is proportional to the NV spin along the NV axis. This term also gives rise to the well-known Einstein-de Haas effect~\cite{Barnett1915,einstein1935can,izumida2022einstein}. 

Refs. \cite{Japha:2022phw, japha2022role} pointed out that the liberation mode gives rise to two significant effects; first of all, the quantum fluctuations in the libration mode depend on the choice of the spin state. The authors showed that the stable states are $|0\rangle, |-1\rangle$ (amongst $|0\rangle, |+1\rangle, |-1\rangle$ states) for the choice of the liberation mode around its minimum, e.g. the libration angle $\beta =0$. However, this choice penalizes the size of the spatial superposition, and typically the maximum superposition size becomes halved. Moreover, the authors chose a simple Stern-Gerlach scheme with a short interferometer duration ($t\sim 100\mu s$) and a relatively small spherical object with a mass ($\sim 10^{-19}$~kg). For these choices, the authors pointed out that even if the NV center is close to the center of mass (C.O.M) the Humpty-Dumpty problem is severe and requires closing the interferometer at specific time dependent on the frequency of the libration mode~\cite{Japha:2022phw, japha2022role}. Although they neglected the diamagnetic-induced contribution in the Hamiltonian $\textbf{B}^2$ term, the spin contrast would not improve by its inclusion for tiny nanorotors.
However, the last assumption must be considered when considering more massive objects, and we do see the mass influencuing the spin contrast in a very interesting manner.

Very recently, in \cite{Zhou:2024pdl}, the authors proposed a completely new scheme. In fact, with one fell of soup, they solved many challenges. They suggested that if we were to initialize the nanorotor with an external rotation along the embedded spin axis (in their analysis it was considered NV spin in a nanodiamond),  then two interesting possibilities emerge: (1) the libration mode $\beta$ gets a new shifted vacuum, and the fluctuations around this new vacuum are highly suppressed, which ameliorates the spin contrast problem, (2) we can allow both $|+1\rangle, |-1\rangle$ states, as shown in an inhomogeneous magnetic field, see~\cite{Marshman:2021wyk} to obtain a large spatial superposition. The superposition size does not get penalized. The rotation imparted in the initialization process depends on the moment of inertia and the external magnetic field at the C.O.M. It is in the range of KHz-MHz (the MHz comes from the EDH effect, particularly to the diamond properties, discussed below). The imparted rotation also ensures that there is no Majorana spin flip, and the Einstein-de Haas effect of the nanorotor is suppressed (we will discuss all these interesting issues in detail).

The purpose of the present paper is to generalize the rotational aspects of nanorotor based on
\cite{Zhou:2024pdl}. First, we wish to generalize our computation to a cylindrical nanorotor. Second, we wish to increase the mass of a nanorotor to $10^{-15}$ kg and study the effect of the Humpty-Dumpty problem. Third, we investigate the role of transversal zero-field splitting in a nanorotor. Typically, this term is negligible, but we will keep it in the dynamics to see how it affects the superposition size. Finally, we wish to study the case where the initial angular states are not in a vacuum but in a thermal state and how it would affect the contrast loss in the SGI setup. We will provide every derivation in detail to illustrate the nanorotor's mechanical properties.

The paper is organized as follows. Section \ref{chapter:NV_Hamiltonian} presents the formalism of the Stern-Gerlach experimental protocol of the spin-embedded defect in a diamagnetic nanorotor, followed by the magnetic field profile that we applied to the Hamiltonian in Section \ref{chapter:magentic_field_profile}. Section \ref{chapter:dynamics_of_euler_angle} provides a formulation of the Euler angle used in this rotational scheme, followed by the Einstein-de Haas effect and how it is mitigated. Section \ref{chapter:spatial_superposition_and_rotational_dynamics} explains the superposition and dynamic rotations in the cylinder case. The last two chapters provide a calculation of the spin contrast in Section \ref{chapter:spin_contrast}, along with the temperature term in Section \ref{chapter:spin_contrast_with_temperature}. Finally, we conclude and summarize our findings in Section \ref{chapter:conclusion}.


\section{Spin-Hamiltonian}
\label{chapter:NV_Hamiltonian}

This section will discuss the Hamiltonian of a levitated nanorotor with an off-center spin-axis (embedded in the crustal) in the presence of an external magnetic field. For numerical purposes, our discussion will closely follow nanodiamond's properties, but as we said, one could apply our methodology to any system with an embedded spin defect with $s=0,\pm 1$ states to manipulate. The nanorotor will be subjected to a sequence of magnetic field gradient pulses and spin-flipping pulses, forming Stern-Gerlach interferometry; for a review, see \cite{Amit2019}. The Hamiltonian of the system is given by \cite{Marshman2022}
\begin{align}
    {H} = \frac{{\mathbf{p}}^2}{2 m}+ \sum_{j=1}^{3} \frac{L^2_j}{2I_j} -\frac{\chi_\rho m}{2 \mu_0} \mathbf{B}^{\mathbf{2}} +\mu {\mathbf{S}} \cdot \mathbf{B} + H_{ZFS},
    \label{eq:hamiltonian_NV_center}
\end{align}

The first term is the kinetic energy in the form of the center of the mass momentum, the second term of $\sum_{j=1}^{3} {L^2_j}/{2I_j}$ is a mechanical angular momentum along the principle axes of rotation $\hat{{n}}_1, \hat{{n}}_2, \hat{{n}}_1$ of the ND $(L_j = \mathbf{L} \cdot \hat{{n}}_j)$, with $I_j$ is the moment of inertia of the nanorotor, the third term is the diamagnetic energy of the nanorotor ($\chi$ is magnetic susceptibility, in this case, diamond which has diamagnetic contribution, $\chi_\rho = -6.2\times10^{-9} \ \text{m}^3/\text{kg}$) in the presence of a magnetic field $\mathbf{B}$. The fourth term represents the energy of the spin in the magnetic field, where $\mu$ is the magnetic moment of an electron. For nanodiamond, {$\mathcal{D}=h\times 2.8 \ \text{MHz}$}. The corresponding $H_{ZFS}$ has been included in this last term as zero-field splitting (ZFS) \cite{Bertrand2020} as
\begin{align}
    {H}_{ZFS} = \mathcal{D} \left[ S_\parallel^2-\frac{S(S+1)}{3}  \right] + E (S_1^2 - S_2^2). \label{eq:hamiltonian_ZFS}
\end{align}
The $\mathcal{D}$ and $E$ parameters refer to axial and transversal ZFS frequency, in which $\mathcal{D}$ is the energy splitting between spin ground state $\ket{0}$ and $\ket{\pm 1}$ and E is the splitting between $\ket{\pm 1}$ due to broken axial symmetry of the NV center caused by the strain effect \cite{Lai2009}. Mathematically, the value of $E$ must be $E \leq \mathcal{D}/3$, experimentally $\mathcal{D}=h\times 2.87$ GHz for the diamond, and the parameter $E$ is different for each unique NV, according to the experimental data \cite{Hoang2016} the value $E\sim 10$ MHz, which assures $E\ll \mathcal{\mathcal{D}}$. In many cases, the $E$ term is often neglected because it represents the transverse component of the magnetic dipole-dipole interaction, which is usually much smaller compared to the axial component, which holds true for diamonds \cite{Bertrand2020}. However, in an Appendix \ref{appendix:E_term}, we will consider a generic value of $E$ to show how it affects the superposition size in a nanorotor (where some specific values will be taken similar to that of a nanodiamond for the purpose of illustration). In this case, the embedded spin on the nanorotor will be projected in the direction of $\hat{n}_s$, so $S_\parallel = \mathbf{S} \cdot \hat{n}_s$, as illustrated in Fig. (\ref{fig:1}.b). 
\begin{figure*}
    \centering
    \includegraphics[width=1\linewidth]{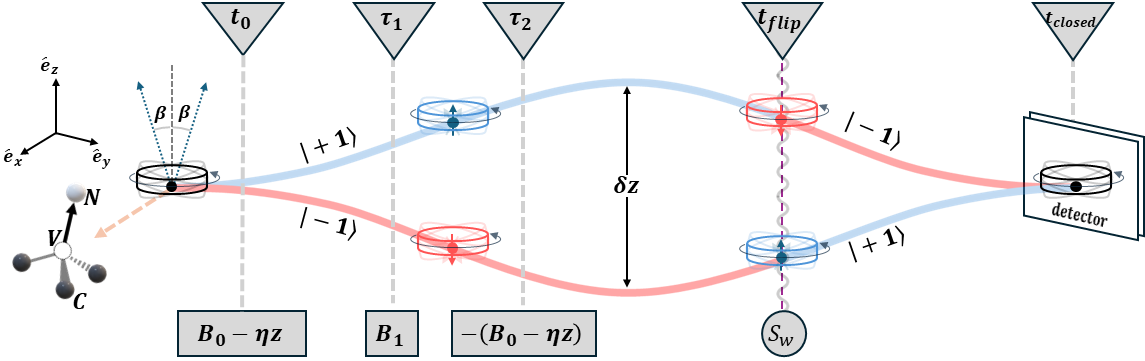}
    \caption{{A side view of a transversal one-loop Stern-Gerlach Interferometer (SGI), z-direction vs time, based on Ref. \cite{Marshman2022}, is shown with modifications for rotation in this case. Three magnetic profiles are used to generate a magnetic field, which first splits ($B_0 - \eta z$) for $0<t<\tau_1$, transitions linearly ($B_1$) for $\tau_1 \leq t \leq \tau_2$, and recombines [$-(B_0 - \eta z)$] for $\tau_2 < t \leq t_{closed}$. The spin is flipped using pulsed microwaves ($S_w$) at time $t=t_{flip}$, moving along the z-axis through the apparatus. The spin quantization axis via the nitrogen-vacancy (NV) nanodiamond is determined by a homogeneous field oriented along the z-axis with $\beta$ as the libration angle. In this research, we set the time values for configuring the magnetic field profile and the time switching function refers to Ref. \cite{Marshman2022}, namely $t_0 = 0 \ \text{s}, \tau_1 = 0.482 \ \text{s}, \tau_2 = 0.514 \ \text{s}, t_{flip} = 0.802 \ \text{s}$ and $t_{closed} = 1.320 \ \text{s}$.}}
    \label{fig:enter-label}
\end{figure*}

Before we proceed further with the case of rotation, we must first determine the equation of motion for the Hamiltonian of the electronic spin $s=0,\pm 1$, as
\begin{align}
    {H}_s = \mu {\mathbf{S}} \cdot \mathbf{B} + \mathcal{D} \left[ S_\parallel^2-\frac{S(S+1)}{3}  \right] +  E (S_1^2 - S_2^2), \label{eq:hamiltonian_spin_only}
\end{align}
where we assumed that the embedded spin is in its ground state, we take spin triplet $S = 1$, so we can rewrite Eq.\eqref{eq:hamiltonian_spin_only} in matrix form
\begin{align}
{H}_s
    & = \begin{pmatrix}
        \mu B_\parallel & \mu \frac{B_\perp}{\sqrt{2}}e^{-i\gamma} & E \\
        \mu \frac{B_\perp}{\sqrt{2}}e^{i\gamma} & -\mathcal{D} & \mu \frac{B_\perp}{\sqrt{2}}e^{-i\gamma} \\
        E & \mu \frac{B_\perp}{\sqrt{2}}e^{i\gamma} & -\mu B_\parallel
    \end{pmatrix}
    + \frac{\mathcal{D}}{3} \mathbb{I}, \label{eq:matrix_form_hamiltonian_with_spin}
\end{align}
where {$B_\parallel$ and $B_\perp $ are respectively the projections of the magnetic field onto the parallel and perpendicular components to the NV axis}, $\gamma$ is the projection angle of the magnetic field on $\hat{n}_1$ and $\hat{n}_2$ plane (see Fig. \ref{fig:1}(a)) and $\mathbb{I}$ is the identity matrix. 

We will simplify the $3\times3$ spin Hamiltonian matrix to a $2\times2$ matrix, retaining only the essential elements using the Feshbach formalism \cite{Feshbach1958,Feshbach1962,Band2022}, we can write the effective Hamiltonian by substituting Eq. \eqref{eq:matrix_form_hamiltonian_with_spin} into (See Appendix \ref{appendix:feshbach} for detail derivation)
\begin{align}
    H_{eff}  =  
    \begin{pmatrix}
        \mu B_\parallel & \epsilon(B_\perp)^* \\
        \epsilon(B_\perp) & -\mu B_\parallel
    \end{pmatrix}
    + \left(\frac{ \mu^2 B_\perp^2}{2\mathcal{D}}+\frac{\mathcal{D}}{3}\right) \mathbb{I}
    \label{eq:H_eff_1}
\end{align}
where $\epsilon(B_\perp) = E+\mu^2 B_\perp^2/(2D)e^{2i\gamma}$, and the $\ast$ is the complex conjugate. We can calculate the eigenvalues of $H_{eff}$ (in our case, {this is the eigenvalues} corresponding to the potential energy of the spin) as
\begin{align}
    V_s & = \pm \sqrt{(\mu B_\parallel)^2+\abs{\epsilon(B_\perp)}^2} +  \left( \frac{\mathcal{D}}{3} + \frac{\mu^2 B_\perp^2}{2\mathcal{D}} \right), \label{eq:V_s}
\end{align}
If we consider $\abs{\mathbf{B}} \ll \mathcal{D}/\mu$, which is practical for experimental parameters, we can state that the term $\mu^2 B_\perp^2/(\mathcal{D}^2) \approx 0$, so $\epsilon \approx E$, also $E\ll \mathcal{D}$. 

Now, we can reduce Eq.\eqref{eq:V_s}, which  becomes
\begin{align}
    V_s & = \pm \sqrt{(\mu B_\parallel)^2+{E}^2} + \frac{\mathcal{D}}{3}, \label{eq:V_s_minimal}
\end{align}
we can calculate the force on the spin using {$\mathbf{F} = -\nabla V_s -\nabla V_{dia}$, where $V_{dia}\propto {\rm B}^2$}, see Eq.~\eqref{eq:hamiltonian_NV_center}.

We are interested in one-dimensional interferometry, which is possible by restricting the motion of the nanorotor along the $z$-axis in a levitating potential; see~\cite{Elahi:2024dbb}.
We get the classical equation of motion for the nanorotor, given by:
\begin{align}
    \Ddot{z} & = \mp  \frac{\mu^2 B_\parallel}{m \sqrt{(\mu B_\parallel)^2+{E}^2}} \frac{\partial B_\parallel}{\partial z} - \frac{\chi_\rho}{2\mu_0} \frac{\partial \mathbf{B}^2}{\partial z}, \label{eq:z_eom}
\end{align}
In order to solve Eq. \eqref{eq:z_eom}, it is essential to establish the magnetic field configuration and the equation governing the rotational degrees of freedom first.


\section{Magnetic field profile for a spatial superposition scheme}
\label{chapter:magentic_field_profile}

The magnetic field profile used in the Hamiltonian eq.\eqref{eq:hamiltonian_NV_center} is based on the Stern-Gerlach scheme employed in Ref. \cite{Marshman2022}, where the magnetic field is described as a function of time as discussed below.
For the applied magnetic field profile at the C.O.M. of the nanorotor:
\begin{align}
    \vec{B}(z,x) = 
    \begin{cases}
        (B_0 - \eta z)\hat{e}_z + \eta x \hat{e}_x, & t < \tau_1 \\
        B_1 \hat{e}_z,              & \tau_1 \leq t \leq \tau_2 \\
        -(B_0 - \eta z)\hat{e}_z - \eta x \hat{e}_x, & t > \tau_2
    \end{cases}
    \label{eq:magnetic_field_profile}
\end{align}
where $B_0$, $B_1$, and $\eta$ are constants with set value $10^{-2}$T, $10^{-4}$T, and $0.45 \times 10^2{\rm T/m}$, {these values are chosen to ensure spin stability and to avoid zero-field regions, as well as to control the trajectory of the particle through the interferometer}. In Ref.~\cite{Marshman2022}, the rotation of the nanorotor is neglected, and the authors assumed that $NV$ is located at the C.O.M, hence $d=0$. Because the nanorotor was assumed to have a spherical symmetry, the NV-axis could be aligned along the z-axis without losing generality. In this paper, we employ a cylindrical geometry, and we also assume that the NV is located  off-centered.
\begin{table*}[t]
\caption{Set up parameters for the nanorotor's mass\footnote{{The chosen mass is within the range of values used in various experiments \cite{jin2024quantum, Segawa2022, Trusheim2013, Neumann2013}}.}, density, magnetic properties of diamond, relevant terms in the spin Hamiltonian, distance of the embedded spin center from the C.O.M., angles, diameter and length of the cylinder.
}
\begin{tabular}{llll}
\hline
\textbf{Parameter}                            & \textbf{Expression}                    & \textbf{Value}                                         & \textbf{unit}     \\ \hline
Mass                              & $m$                           & $ 5 \times 10^{-18} - 10^{-15}$                               & kg       \\
Density                           & $\rho_D$                        & $3.5 \times 10^3$                             & $\text{kg}/\text{m}^3$ \\
Magnetic susceptibility of Diamond    & $\chi_\rho$                   & $-6.2 \times 10^{-9}$                         & $\text{m}^3/\text{kg}$ \\
Free Space Permeability               & $\mu_0$                       & $1.257 \times 10^{-6}$                        & H/m      \\
Transversal NV Zero-field splitting  & $\mathcal{D}$                      & $h\times 2.87 \times 10^9$                    & Hz       \\
Longitudinal NV Zero-field splitting & $0 \leq E \leq \mathcal{D}_{NV}/3$ & $0 \leq E \leq h\times 0.95 \times 10^9$ & Hz       \\
 Spin off-center distance               & $d$                           & $10 \times 10^{-9}$                           & m        \\
Angle between spin-axis and d          & $\alpha'$                     & $\pi/6$                                       & rad      \\
Magnetic Moment of spin                & $\mu$                         & $h\times 2.8\times 10^{10}$                     & Hz/T     \\
Initial $\beta$ angle    & $\beta_0$                    & 0.01                            & rad    \\
Ratio of diameter-to-height of cylinder                   & $D/L$                           & $0.1 \ (\text{long cylinder}), 1 \ (\text{normal}), \ \text{and} \ 10\ (\text{disk-shaped})$          & -                \\ \hline
\end{tabular}
\label{table:1}
\end{table*}
The external magnetic field applied along the $\hat{e}_z$ direction; see Fig.~\ref{fig:1}. Although our magnetic field profile also provides a component in the $\hat{e}_x$ direction, we restrict the superposition along $\hat{e}_z$. This is possible by restricting the motion of the particle along the $\hat{e}_z$ axis by creating an appropriate diamagnetic trap~\footnote{Creating a trap goes beyond the scope of the current paper, here we will assume that a diamagnetic trap can be constructed with a very steep potential in the $\hat{e}_x$ direction, while the $\hat{e}_z$ direction is a flat one, where the superposition can be created. This assumption was also made to analyse the libration mode~\cite{Japha:2022phw}.}. Therefore, we always set $x = 0$ for computation.
The switching mechanism in the magnetic profile for Eq. \eqref{eq:magnetic_field_profile} has been explained in Ref. \cite{Marshman2022}, and we summarise it below.
\begin{enumerate}
    \item At $t<\tau_1$, $\vec{B}$ is given by $(B_0 - \eta z)\hat{e}_z$. {This field is used to separate interferometer paths based on the spin of the particle. The magnetic field gradient causes nanodiamond with different spins to experience different forces, resulting in path separation.}
    
    \item For $\tau_1 \leq t \leq \tau_2$, $\vec{B}$ is switching adiabatically to $B_1 \hat{e}_z$. {this stage aim to keep the nanodiamon paths separated without further increasing or decreasing the separation. This constant field avoids further changes in the momentum of the nanodiamons.}
    
    \item For $t > \tau_2$, $\vec{B}$ is switching adiabatically to $-(B_0 - \eta z)\hat{e}_z$. {This stage aims to recombine the interferometer paths by directing the nanodiamond back to their initial positions, allowing interference to occur.} 
    
    \item at $t = t_{flip}$, $\vec{B}$ is still the same as at point 3; however, the spin state of the embedded spin, $s$, is switched ({$\{\ket{\pm 1} \rightarrow \ket{\mp 1}\}$}) to close the superposition using fast pulsed microwave. {More importantly, implementing $t_{flip}$ in this scheme aims to achieve a smooth transition, avoiding sudden changes in the magnetic field that can cause spin instability or loss of quantum coherence. Thus, the timing of the flip is crucial to ensure that the interferometer paths can be precisely controlled.}
    
    \item At $t = t_{closed}$, the two wave functions (left and right arms of the interferometer) overlap in position and momentum bases.
    
\end{enumerate}

During $\tau_1 \leq t \leq \tau_2$, the nanorotor experiences a sufficiently large magnetic field to avoid Majorana spin flip; see \cite{Marshman2022}. {Another critical aspect to consider is the interferometer time scale, which is proportional to the inverse of the magnetic field gradient $\eta^{-1}$ or in range $\sim 0.01 - 1$ s \cite{Zhou:2024pdl}}. To close the two SGI arms, a fast microwave pulse with the energy difference between two spin states allows the transition from spin up ($\ket{+1}$) to spin down ($\ket{-1}$) and vice versa at time $t_{flip}$ and closed superposition at $t_{closed}$.  
In our case, we set the values $\tau_1 =0.482~{\rm s},~\tau_2=0.514$ s and $t_{flip}=0.802~{\rm s},~ t_{closed}=1.32$ s.

We denote $B_\parallel = B_z \cos{\beta}$ defined as a magnetic field projection in the NV-axis ($\hat{n}_s$) direction, in the first term of Eq.\eqref{eq:z_eom}. Using the definition of a magnetic field profile Eq.\eqref{eq:magnetic_field_profile}, we can re-write the C.O.M equation of motion along the z-component:
\begin{align}
    \Ddot{z} = &  - \frac{s\mu^2 \Tilde{\eta}^2(t) (z-Z_0) \cos^2{\beta}}{m \sqrt{\mu^2 \Tilde{\eta}^2(t) (z-Z_0)^2\cos^2{\beta}+{E}^2}} \nonumber \\
    &- \frac{\chi_\rho\Tilde{\eta}^2(t)}{\mu_0} (z-Z_0),
    \label{eq:Ddot_z_general}
\end{align}
where $Z_0 = B_0/\eta$, $s=\{-1,0,+1\}$ represents the spin state, and $\Tilde{\eta}(t)$ is the magnetic field gradients which depend on time slices following the Stern-Gerlach scheme~\cite{Marshman2022}.
\begin{align}
    \Tilde{\eta}(t) = 
    \begin{cases}
        -\eta, & t < \tau_1 \\
        0,              & \tau_1 \leq t \leq \tau_2 \\
        \eta, & t > \tau_2
    \end{cases}
    \label{eq:magnetic_gradient}
\end{align}
We consider the strain frequency, $E\approx 0$ (typically, it is small in the nanodiamond,  see~\cite{Hoang2016,Bertrand2020}
). In such a case,  the spatial equation of motion becomes:
\begin{align}
    \Ddot{z} = - \frac{s \mu \Tilde{\eta}(t)}{m} \cos{\beta}  - \frac{\chi_\rho \Tilde{\eta}^2(t)}{\mu_0} (z-Z_0)\,.
    \label{eq:Ddot_z}
\end{align}
We also analyze the non-zero value of $E$ in the Appendix \ref{appendix:E_term}; see Fig.~\ref{fig:zero_and_-1}.


\section{Euler Angles in the cylindrical nanorotor}
\label{chapter:dynamics_of_euler_angle}
\begin{figure*}
    \centering
    \includegraphics[width=1\linewidth]{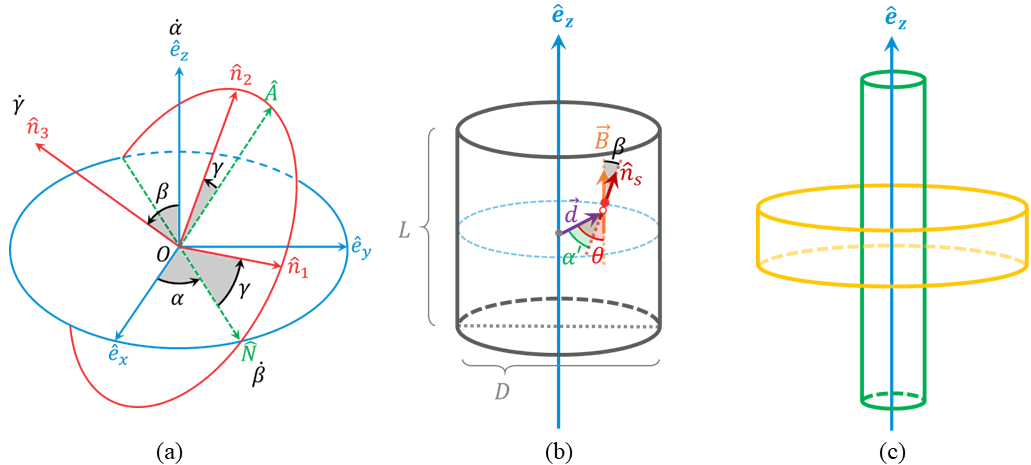}
    \caption{{(\textit{a}) Euler classical coordinate with angle ($\alpha, \beta, \gamma $), blue line define as fixed coordinate system ($\hat{e}_x, \hat{e}_y, \hat{e}_z$), red line as rotated coordinate ($\hat{n}_1, \hat{n}_2, \hat{n}_3$), and green line as line of nodes ($N$). The sketch (b) in the right side are the cylinder nanorotor with off-centered spin fixed in the crystal, where the spin axis $\hat{n}_s$ is parallel to $\hat{n}_3$. $\beta$ is the angle between spin-axis and z-axis, $\alpha'$ is fix angle between NV-axis and the off-center vector $\vec{d}$, and $\theta$ is the angle between $\vec{d}$ and z-axis, with the relation $\beta = \theta - \alpha'$. The external magnetic field is applied along $\hat{e}_z$ direction (Although our magnetic field profile also provides a component in the $\hat{e}_x$ direction, we restrict the superposition to occur along $\hat{e}_z$. Therefore, we always set $x = 0$. Since the nanodiamond is cylindrical in shape, we consider its form with three interrelated parameters: mass, diameter, and height of the cylinder, in sketch (c) We set the ratio of diameter-to-height value ($D/L$) in such a way that three alternative shapes of the cylinder are obtained in same mass, which are normal cylinder (grey) at $D/L\approx1$, disk-shaped (yellow) at $D/L\approx 10$, and long cylinder (green) at $D/L\approx 0.1$. We limit the discussion of rotational dynamics and contrast only to the cases of normal cylinders and disk-shaped objects, as these forms are more stable compared to long-cylinders (See Appendix \ref{appendix:long_cylinder} for the discussion on the dynamics and contrast of long-cylinders).}}
    \label{fig:1}
\end{figure*}

Before we investigate the rotation effect of the cylinder, we first discuss the effect of an embedded spin on free rotation and torque. In the comoving frame of a rigid body, the coordinates are given by: $\hat{n}_1$, $\hat{n}_2$, $\hat{n}_3$. Relative to the laboratory frame, the coordinates are $\hat{e}_x, \hat{e}_y, \hat{e}_z$. These two coordinates are typically described through three successive elementary rotations using Euler angles $\alpha$, $\beta$, and $\gamma$, see Fig.~\ref{fig:1}(a). In this figure, $\hat{n}_3$ is considered the cylinder's symmetry axis, so the moment of inertia is set to be: $I_1 = I_2 = I \neq I_3$. The rotation of the body results in three angular velocities:
\begin{align}
    \boldsymbol{\omega} = \dot{\alpha} \hat{e}_z + \dot{\beta} \hat{N} + \dot{\gamma} \hat{n}_3,
    \label{eq:omega_vector}
\end{align}
where dot denotes time derivative, $d/dt$, and 
$\hat{N}$ is the unit vector along the line  ON. Along with the auxiliary vector $\hat{N}$, we will need a vector $\hat{A}$, which lies on {the plane spanned by $\hat{n}_1$ and $\hat{n}_2$, and is perpendicular to both $\hat{N}$ and $\hat{n}_3$}. Hence, 
$\hat N,~\hat e_z,~\hat A$ are given by:
\begin{subequations}
\begin{align}
    \hat{N} & = \cos{\gamma} \hat{n}_1 - \sin{\gamma} \hat{n}_2, \\
    \hat{e}_z & =  \sin{\beta} \hat{A} + \cos{\beta} \hat{n}_3, \label{eq:natural_vector} \\
    \hat{A} & = \sin{\gamma} \hat{n}_1 + \cos{\gamma} \hat{n}_2,
\end{align}
\end{subequations}
we substitute Eq.\eqref{eq:natural_vector} into Eq.\eqref{eq:omega_vector}, and we have, $\boldsymbol{\omega}$, that projected onto the comoving frame (attached to the cylinder), given by:
\begin{align}
    \boldsymbol{\omega} =
    \begin{bmatrix}
        \omega_1 \\
        \omega_2 \\
        \omega_3
    \end{bmatrix} =
    \begin{bmatrix}
        \dot{\alpha} \sin{\beta} \sin{\gamma} + \dot{\beta} \cos{\gamma} \\
        \dot{\alpha} \sin{\beta} \cos{\gamma} - \dot{\beta} \sin{\gamma} \\
        \dot{\alpha} \cos{\beta} + \dot{\gamma}
    \end{bmatrix},
    \label{eq:omega123}
\end{align}
and if we project onto the laboratory frame (defined by $x,y,z$), we have:
\begin{align}
    \boldsymbol{\omega} =
    \begin{bmatrix}
        \omega_x \\
        \omega_y \\
        \omega_z
    \end{bmatrix} =
    \begin{bmatrix}
        \dot{\gamma} \sin{\beta} \sin{\alpha} + \dot{\beta} \cos{\alpha} \\
        \dot{\gamma} -\sin{\beta} \cos{\alpha} + \dot{\beta} \sin{\alpha} \\
        \dot{\gamma} \cos{\beta} + \dot{\alpha}
    \end{bmatrix},
\end{align}
The complete relationship between the two set of vectors $\hat{n}_1$, $\hat{n}_2$, $\hat{n}_3$ and { $\hat{e}_x, \hat{e}_y, \hat{e}_z$} in terms of the Euler angles is
now given by the matrix transformation:
\begin{widetext}
    \begin{align}
        \begin{bmatrix}
            \hat{n}_1 \\
            \hat{n}_2 \\
            \hat{n}_3 
        \end{bmatrix}
         =
        \begin{bmatrix}
            \cos{\alpha} \cos{\gamma} - \cos{\beta}\sin{\alpha}\sin{\gamma} & \sin{\alpha} \cos{\gamma}+ \cos{\beta}\cos{\alpha}\sin{\gamma} & \sin{\beta} \sin{\gamma} \\
            -\cos{\alpha} \sin{\gamma} - \cos{\beta}\sin{\alpha}\cos{\gamma} & -\sin{\alpha} \sin{\gamma}+ \cos{\beta}\cos{\alpha}\cos{\gamma} & \sin{\beta} \cos{\gamma} \\
            \sin{\beta}\sin{\alpha} & - \sin{\beta}\cos{\alpha}  & \cos{\beta} 
        \end{bmatrix}
        \begin{bmatrix}
            \hat{e}_x \\
            \hat{e}_y \\
            \hat{e}_z 
        \end{bmatrix}
        \label{eq:transformation}
    \end{align}
\end{widetext}
Here, we considered $\hat{e}_z$ as a fixed direction along the total spatial momentum vector.

We can calculate the rotational kinetic energy ($T_{rot} = \frac{1}{2} \sum_{i = 1}^3 \ {L}^2_i/I_i$) of the cylinder in terms of the Euler angles using the Eq.~\eqref{eq:omega123} by using the relation $\omega_i = L_i/I_i$, we have\cite{Zhou:2024pdl,Japha:2022phw}
\begin{align}
    T_{rot} = \frac{1}{2} I\dot{\alpha}^2 \sin^2{\beta} + \frac{1}{2} I \dot{\beta}^2 + \frac{1}{2} I_3 ( \dot{\alpha} \cos{\beta} + \dot{\gamma})^2.
    \label{eq:kinetic_energy_rotation}
\end{align}
To complete the Hamiltonian of a rigid rotor, we need to consider Eq.~\eqref{eq:matrix_form_hamiltonian_with_spin}, especially the $\mathcal{D}$-term that will generate the  Zeeman and the Einstein-de Haas terms.

\subsection{Einstein de Haas term: interplay between NV spin and angular momentum}

Note that Eq.\eqref{eq:hamiltonian_spin_only} composed of the Zeeman term, $\mu \textbf{S} \cdot \textbf{B}$, and the $\mathcal{D}$ parameter, $\mathcal{D}S_\parallel$. However, once the rotation is introduced to the nanorotor, then there will be an additional term involved, which we wish to investigate.

During the nanorotor's rotation, the spin vector of the NV center is assumed to remain aligned with the NV axis. The mechanical rotation of the nanorotor will thus cause the magnetic moment of the NV spin and the spin angular momentum to rotate. Under these conditions, the spin of the NV center will experience two types of torques due to the rotation of the nanorotor.

\begin{enumerate}
    \item \textbf{The Zeeman term}, $\mu (\mathbf{S}\cdot\mathbf{B})$: This term occurs when the magnetic moment of the NV spin interacts with the external magnetic field. The Zeeman effect causes the energy levels of the embedded spin to split according to the orientation of its magnetic moment relative to the magnetic field.
    
    \item \textbf{The Einstein-de Haas term}:
    This is a physical phenomenon in which a rigid object rotates due to a change in the magnetic moment of the object. This effect is a consequence of the conservation of angular momentum, see \cite{barnett1915magnetization,einstein1915experimental}. In this context, the rotation of the nanorotor can be considered a response to changes in the angular momentum of the spin, given by: $\dot{\mathbf{S}} = \mathbf{S} \cross \boldsymbol{\omega}= \mathbf{S} \cross \mathbf{L}/\mathbf{I}$, highlighting the interconnection between the spin motion and the rotational/angular momentum $\mathbf{L}$, see the derivation below. 
    
\end{enumerate}

The interaction between the torques (due to the external magnetic field) resulting from the Zeeman term creates complex dynamics in the system, and there will be an additional coupling between $\mathbf{S}$ and $\mathbf{L}$. The mathematical expression for these two terms can be shown in the evolution of spin and angular momentum of nanorotor's rotation, (recall Eq. \eqref{eq:Si_time_derivative_A} and Eq.\eqref{eq:Li_time_derivative_A} in Appendix \ref{appendix:A}):
\begin{align}
    \dot{S}_i & = \frac{i}{\hbar} [\mathcal{D}S_\parallel^2,S_i]+ \frac{\epsilon_{ijk}}{\hbar} \left(\mu S_j B_k + \hbar S_j \frac{L_k}{I_k} \right), \label{eq:Si_time_derivative} \\
    \dot{L}_i & = \epsilon_{ijk} \left(\mu B_j S_k + {\hbar} S_j \frac{L_k}{I_k}+L_j\frac{L_k}{I_k} \right), \label{eq:Li_time_derivative}
\end{align}
where $(i,j,k)$ take values $1,2,3$. Note that in the last term of Eq.\eqref{eq:Si_time_derivative_A}, we have a term, e.g. $\Gamma \sim \hbar/I$. However, thanks to a large moment of inertia $I\sim 10^{-32} {\rm \text{kgm}^2}$, we have $\Gamma \sim 10^{-12}$ Hz, which is much smaller compared to $\mathcal{D}\sim 10^9 \ \text{Hz}$, so this term can be neglected in Eq.\eqref{eq:Si_time_derivative_A}.

Note that in the evolution of $L_{i}$, we have the
Zeeman torque (first term), the Einstein-de Haas term (second term), and the last part is the rotational kinetic term. The most interesting thing is that even if no magnetic field ($B=0$) is applied to the nanorotor, the NV spin will still experience the term ${\hbar} S_j {L_k}/{I_k}$ via the Einstein-de Haas contribution due to the presence of the $\mathcal{D}$-term in the crystal, see Appendix ~\ref{appendix:A}.

Now, we can define the effective spin Hamiltonian to describe the spin evolution in the rotating frame as
\begin{align}
    \widetilde{H}_{s} & = \mu \mathbf{S} \cdot \mathbf{B} + \mathcal{D}\left[S^2_\parallel -\frac{S(S+1)}{3}\right] - \hbar \mathbf{S}\cdot\frac{\mathbf{L}}{\mathbf{I}},
    \label{eq:H_with_edh}
\end{align}
can be verified that the Hamiltonian, $\widetilde H_s$, satisfies equation $\dot{S}_i = i{[\widetilde{H}_s, S_i] }/{ \hbar}$ (see Appendix \ref{appendix:A}). Furthermore, Equation \eqref{eq:Si_time_derivative} is satisfied by the same Hamiltonian $\widetilde H_s$. We can split Equation \eqref{eq:H_with_edh} into two parts as follows::
\begin{equation}
    \widetilde{H}_s = \widetilde{H}^{(1)}_s + \widetilde{H}^{(2)}_s
\end{equation}
where $\widetilde {H}^{(1)}_s$ corresponds to the Zeeman and Zero-Field Splitting term represented in a matrix form of Eq. \eqref{eq:matrix_form_hamiltonian_with_spin}:
\begin{align}
    \widetilde{H}^{(1)}_s & = \mu \mathbf{S} \cdot \mathbf{B} + \mathcal{D}\left[S^2_\parallel -\frac{S(S+1)}{3}\right] \nonumber \\
    & =  \begin{pmatrix}
        \mu B_\parallel & \mu \frac{B_\perp}{\sqrt{2}}e^{-i\gamma} & 0 \\
        \mu \frac{B_\perp}{\sqrt{2}}e^{i\gamma} & -\mathcal{D} & \mu \frac{B_\perp}{\sqrt{2}}e^{-i\gamma} \\
        0 & \mu \frac{B_\perp}{\sqrt{2}}e^{i\gamma} & -\mu B_\parallel
    \end{pmatrix}
    + \frac{\mathcal{D}}{3} \mathbb{I},
\end{align}
and $\widetilde{H}^{(2)}_s$ incorporates the Einstein-de Haas term. For cylindrical nanorotor, we can express the matrix as 
\begin{align}
    \widetilde H^{(2)}_s & = -\hbar \ \mathbf{S}\cdot \frac{\mathbf{L}}{\mathbf{I}} \nonumber \\
    & = -\begin{pmatrix}
        \hbar \frac{L_3}{I_3} & \frac{\hbar}{I} \frac{L_1 -iL_2}{\sqrt{2}} & 0 \\
        \frac{\hbar}{I} \frac{L_1 +iL_2}{\sqrt{2}} & 0 & \frac{\hbar}{I} \frac{L_1 -iL_2}{\sqrt{2}} \\
        0 & \frac{\hbar}{I} \frac{L_1 +iL_2}{\sqrt{2}} & -\hbar \frac{L_3}{I_3}
    \end{pmatrix}\,.
    \label{eq:Hs_2}
\end{align}
However, in this paper, we will see that our parameter space prefers $\widetilde H^{(1)}_s \gg \widetilde H^{(2)}_s$ for the external magnetic field required to create spatial superposition.


\subsection{Gyroscopic Stability model for matter-wave interferometery}

As explained in the Zhou-Bose-Mazumdar for a spherical nanocrystal, large spatial superposition can be ensured along with the stability of $|-1\rangle$ and $|+1\rangle $ states, provided we impart an initial rotation along the direction of  $\hat{n}_3$, which aligns with the $\hat{e}_z$ as close as possible~\cite{Zhou:2024pdl}. 

Fundamentally, there will be a small initial angle, denoted here by $\beta_0$, between $\hat{n}_3$ and $\hat{e}_z$. We will call it a libration angle. According to the protocol of Zhou-Bose-Mazumdar, we will introduce initial $\omega_0$. we will be operating at a frequency smaller than the Einstein-de Haas contribution (see the discussion below), e.g. $\mathbf{L}/\mathbf{I} = \boldsymbol{\omega}$, so that 
\begin{align}
    \widetilde{H}^{(2)}_s 
    \approx -\begin{pmatrix}
        \hbar \omega_3 & {\hbar} \frac{\omega_1 -i\omega_2}{\sqrt{2}} & 0 \\
        {\hbar} \frac{\omega_1 +i\omega_2}{\sqrt{2}} & 0 & {\hbar} \frac{\omega_1 -i\omega_2}{\sqrt{2}} \\
        0 & {\hbar} \frac{\omega_1 +i\omega_2}{\sqrt{2}} & -\hbar \omega_3
    \end{pmatrix}, 
\end{align}
From Eq.\eqref{eq:omega123}, we set $\omega_0$ as an initial rotation solely along the rotation angle $\gamma$, so that $\dot{\gamma}(0) = \omega_0$ and $\dot{\alpha}(0) = \dot{\beta}(0) = 0$, and for the initial angles  $\beta(0) =\beta_0$ and $\alpha(0) = \gamma(0) = 0$, so we have
\begin{align}
    \boldsymbol{\omega}_0 =
    \begin{bmatrix}
        0 \\
        0 \\
        \omega_0
    \end{bmatrix},
    \label{eq:omega_0-123}
\end{align}
While substituting Eq.\eqref{eq:omega_0-123} into Eq.\eqref{eq:Hs_2}, we obtain:
\begin{align}
    \widetilde H^{(2)}_s 
    \approx -\begin{pmatrix}
        \hbar \omega_0  & 0 & 0 \\
        0 & 0 & 0 \\
        0 & 0 & -\hbar \omega_0
    \end{pmatrix}, 
\end{align}
Now we have two matrices $\widetilde{H}^{(1)}_s$ and $\widetilde H_s^{(2)}$, hence the effective matrix is given by $\widetilde H^{eff}_s=\widetilde H^{(1)}_s+\widetilde H^{(2)}_s$:
\begin{widetext}
\begin{align}
    \widetilde{H}^{eff}_s = &  
    \begin{pmatrix}
        \mu B_z \cos{\beta_0} - \hbar\omega_0  & \mu \frac{B_z \sin{\beta_0}}{\sqrt{2}}e^{-i\omega_0 t} & 0 \\
        \mu \frac{B_z\sin{\beta_0}}{\sqrt{2}}e^{i\omega_0 t} & -\mathcal{D} & \mu \frac{B_z\sin{\beta_0}}{\sqrt{2}}e^{-i\omega_0 t} \\
        0 & \mu \frac{B_z \sin{\beta_0}}{\sqrt{2}}e^{i\omega_0 t} & -\mu B_z \cos{\beta_0} + \hbar\omega_0
    \end{pmatrix} 
     + \frac{\mathcal{D}}{3} \mathbb{I},
    \label{effective-spin}
\end{align}
\end{widetext}

\subsection{Effective Spin Hamiltonian of a nanorotor}

We can use the same Feshbach technique to reduce Eqs. \eqref{eq:H_with_edh} and \eqref{effective-spin} based on the derivation of Eq. \eqref{eq:H_eff_1}, such that the effective Hamiltonian Eq.\eqref{effective-spin} becomes
\begin{align}
    \widetilde H_{s}^{eff}  =  
    \begin{pmatrix}
        \Delta_+ & W^* \\
        W & \Delta_-
    \end{pmatrix}
    + \frac{\mathcal{\mathcal{D}}}{3} \mathbb{I},
    \label{Eff-Ham}
\end{align}
where 
\begin{eqnarray}
&&\Delta_\pm = \pm  \mu B_z \cos{\beta_0} \mp \hbar \omega_0 \,,\\
&& W = \frac{\mu^2 B_z^2 \sin^2{\beta_0}}{2\mathcal{D}}e^{2i\omega_0 t}
\end{eqnarray}
Two cases arise now, namely Majorana spin-flip and Rabi oscillation, which we will discuss briefly:
\begin{enumerate} 
    \item {\bf Majorana Spin Flip}: If $\hbar\omega_0 \approx \mu B_z$, a Majorana spin flip phenomenon will occur from $\ket{+1}$ to $\ket{-1}$ and vice versa. We impose conditions such as $\omega_0 \ll \mu B_z/\hbar$ to mitigate this issue. The Larmour frequency of an electron is given by $\omega= eB/m_e = 2\mu B/\hbar$.

    \item {\bf Spin-resonant transition}: If $\hbar\omega_0 \approx \mathcal{D} \pm \mu B_z$, $\ket{+1}$ or $\ket{-1}$ will degenerate with the $\ket{0}$ state, causing the Rabi oscillation between these two states. To address this issue, we impose the condition $\omega_0 \ll (\mathcal{D} \pm \mu B_z)/\hbar $. Typically, in any experiment $\mathcal{D}\gg \mu B_z$.

\end{enumerate}

Consequently, to avoid Majorana flip and Rabi oscillation in the embedded spin state in a nanorotor, we set the following condition on our initial rotation of the nanorotor along the direction of $B_z$ before we begin the spatial interferometer. The constraint on $\omega_0$ ensures that we are always below the frequency of the Einstein-de Haas contribution.

\begin{equation}
\omega_0\ll \frac{\mu B_z}{\hbar}\, ,\  {\rm and}\ \omega_0\ll \frac{\mathcal{D}\pm\mu B_z}{\hbar}\label{eq:EdH_constrain} .
\end{equation}
With these constraints in mind, we can now focus on the Zeeman term in Eq.~\eqref{eq:hamiltonian_spin_only}.

\section{ Spatial superposition combined with rotational dynamics}
\label{chapter:spatial_superposition_and_rotational_dynamics}

Refs.\cite{Zhou:2024pdl} and \cite{Japha:2022phw} successfully investigated the rotation of a spherical nanodiamond with an 
off-centered NV system. The fundamental difference is that, in the case of gyroscopic stability, Ref.~\cite{Zhou:2024pdl} introduced an initial angular velocity along the axis of rotation that is aligned with the direction of the embedded electronic spin. This has a distinct advantage over
\cite{Japha:2022phw} because it addresses the spin contrast successfully, and the scheme enhances the spatial superposition size, which is compatible with \cite{Marshman:2021wyk}. We discussed  the results of \cite{Japha:2022phw} briefly in Appendix \ref{appendix:A}. Inspired by \cite{Zhou:2024pdl}, we will consider a cylinder with $I_1 = I_2 = I \neq I_3$. {This inhomogeneous moment of inertia results in three different cases for the shape of the cylinder: long, normal and disk-shaped cylinders, and it affects the rotational dynamics as shown in Figs. \ref{fig:1} (b) and (c)}. To create spatial superposition, we can envision the experimental steps as follows:

\begin{enumerate} 
    \item Initially, we suspended a cylindrical nanorotor and applied circularly polarised light to set the nanorotor rotating with an angular speed $\omega_0$ along the $\hat{e}_z$-axis. This approach aligns with the experimental method utilised in Ref. \cite{jin2024quantum} for spherical nanorotors. The employment of circularly polarised light ensures consistent and uniform rotation, which is crucial to achieving the target angular speed.
   
    \item Subsequently, we configure the electronic spin state embedded in a nanorotor into a superposition state represented as $\ket{S} = (\ket{+1} + \ket{-1})/{\sqrt{2}}$. This spin-state superposition (internal degree of freedom) is pivotal for enabling further manipulation and observation of the quantum properties of the nanrotor, and creating spatial superposition.
   
    \item Lastly, at the moment $t = 0$, we introduce a magnetic field $B_z$. The direction $\hat{e}_z$ should align as closely as possible to the nanorotor's initial rotation axis $\hat{n}_3$. However, our experimental arrangement can accommodate a minor initial angle $\beta_0$ between $\hat{n}_3$ and $\hat{e}_z$.
\end{enumerate}

In the next subsection, we will start with the rotational dynamics of the cylindrical nanorotor, then compare the cases with and without the initial rotation $\omega_0$, and the evolution of $\alpha, \beta, \ \text{and} \ \gamma$ during the spatial superposition.


\subsection{Rotation and center of mass dynamics}

We consider a cylindrical nanorotor of mass $m$, height $L$, and radius $R$, where {$\rho = 3.5 \times 10^3$ $\text{kg}/\text{m}^3$}, which is that of a diamond. The electronic spin is embedded in the nanorotor at a distance $d$ from the C.O.M. The embedded spin-axis is fixed in an angular direction $\alpha'$ from the vector $\vec{d}$ as shown in Fig. \ref{fig:1}(b), and see Table \ref{table:1} for the numerical values. 

To simplify, we set the embedded spin-axis $\hat{n}_s$ parallel to $\hat{n}_3$. The angle of NV $\beta= \theta - \alpha'$ and the definition of $\vec{d}$ and $\hat{n}_s$ are give by:
\begin{align}
    \vec{d} & = d (\sin{\theta} \hat{e}_y +  \cos{\theta} \hat{e}_z), \label{eq:d} \\
    \hat{n}_s & = \sin{\beta} \hat{e}_y + \cos{\beta} \hat{e}_z \label{eq:n_3}\, .
\end{align}
The total Hamiltonian of the nanorotor with defect situated off-centered is given by total Hamiltonian $H_t$:
\begin{align}
    H_t = T_{rot} + \widetilde H^{eff}_{s},
\end{align}
where $T_{rot}$ described as Euler angle in eq. \eqref{eq:kinetic_energy_rotation}. Then, We can write kinetic as a form of canonical momentum, $(p_\alpha, p_\beta, p_\gamma)$, are given by:
\begin{subequations}
\begin{align}
    p_\alpha & = \frac{\partial T_{rot}}{\partial \dot{\alpha}} =  I\dot{\alpha} \sin^2{\beta} + p_\gamma \cos{\beta}, \\
    p_\beta & = \frac{\partial T_{rot}}{\partial \dot{\beta}} = I \dot{\beta}, \\
    p_\gamma & = \frac{\partial T_{rot}}{\partial \dot{\gamma}} = I_3 ( \dot{\alpha} \cos{\beta} + \dot{\gamma}),
\end{align}    
\end{subequations}
If we reformulate $\dot{\alpha}, \dot{\beta}, \dot{\gamma}$ in form of $(p_\alpha, p_\beta, p_\gamma)$, then in terms of $\dot \alpha, \dot \beta, \dot \gamma$, we have:
\begin{subequations}
\begin{align}
    \dot{\alpha} & = \frac{p_\alpha - p_\gamma \cos{\beta}}{I\sin^2{\beta}}, \\
    \dot{\beta} & = \frac{p_\beta}{I}, \\
    \dot{\gamma} & = \frac{p_\gamma}{I_3} - \left( \frac{p_\alpha - p_\gamma \cos{\beta}}{I\sin^2{\beta}} \right)\cos{\beta}
\end{align}    
\end{subequations}
Now, we can rewrite Eq. \eqref{eq:kinetic_energy_rotation}, $T_{rot}$, as function of $p_\alpha, p_\beta, p_\gamma$:
\begin{align}
    T_{rot} = \frac{p_\beta^2}{2I} + \frac{p_\gamma^2}{2I_3} +  \frac{(p_\alpha - p_\gamma \cos{\beta})^2}{2I\sin^2{\beta}}, 
\end{align}
so we have the Hamiltonian
\begin{align}
    H_t = \frac{p_\beta^2}{2I} + \frac{p_\gamma^2}{2I_3} + \frac{(p_\alpha - p_\gamma \cos{\beta})^2}{2I\sin^2{\beta}} + \widetilde H^{eff}_{s}\,,
    \label{spin-Ham0}
\end{align}
where we have added the effective spin part of the Hamiltonian from Eq.~\eqref{Eff-Ham}.

To find the equations of motion of the spin part of the Lagrangian,  where $\psi =\{ \alpha, \beta, \gamma \}$:
\begin{align}
    p_\psi = \frac{\partial \mathcal{L_s}}{\partial \dot{\psi}}\, , \quad
    \frac{dp_\psi}{dt} = \frac{\partial \mathcal{L_s}}{\partial \psi}, 
\end{align}
Because the Lagrangian depends on angle $\beta$ and $\partial \mathcal{L_s}/ \partial\dot{\alpha} = \partial \mathcal{L_s}/ \partial\dot{\gamma} = \text{const} $, it follows that $p_\alpha = L_z$ and $p_\gamma = L_3$, meaning that $\alpha$ and $\gamma$ are cyclic coordinates that are absent from the Hamiltonian, which is why there is conservation of their conjugate momenta. Hence, we have to calculate the equation of motion only for $\beta$, given by:
\begin{align}  
    \Ddot{\beta}_{rot} & = \frac{(p_\alpha - p_\gamma \cos{\beta})( p_\alpha \cos{\beta} -p_\gamma)}{I^2 \sin^3{\beta}}
    \label{eq:beta_rot}
\end{align}
The off-centered embedded electronic spin defect can be illustrated in Fig. \ref{fig:1}(b). The spin Hamiltonian of written as follows ~\footnote{Note that here we are discussing only the embedded spin interaction with the external magnetic field. However, in the equations of motion, while creating superposition, we have included the effect of the diamagnetic-induced magnetic term, see Eq.\eqref{eq:z_eom}.}
\begin{align}
    \widetilde H^{eff}_{s} \approx \mu \mathbf{S} \cdot \mathbf{B}(\vec{z} + \vec{d}) + H_{ZFS},
    \label{spin-Ham1}
\end{align}
where the dimensionless $\mathbf{S}$ align along $\hat{n}_3$ at an angle $\beta$ with $\mathbf{B}$ at an off-center distance $\vec{d}$ with respect to the C.O.M. The distance $\vec{d}$ is directed with an angle $\theta$ with respect to the z-axis. The angle $\beta$ is known as the libration angle or libration mode, as discussed above.

We use the magnetic field profile at the C.O.M of nanorotor given by Eq. \eqref{eq:magnetic_field_profile}. The NV spin $\hat{n}_s$, given by $\mathbf{S} = s \hat{n}_s$, takes values, $s = \{ +1, -1, 0\}$. The angle $\alpha'$ is fixed between $\vec{d}$ and $\hat{n}_s$, where we set the NV-spin axis parallel to $\hat{n}_3$ ($\hat{n}_s =  \hat{n}_3$). With our definition of $\vec{d}$ in Eq. \eqref{eq:d} and magnetic field profile given by \eqref{eq:magnetic_field_profile}, we can define the magnetic field strength at the NV off-centered location, as
\begin{align}
    B_{nv}(\vec{z}+\vec{d}) = B_z + \tilde{\eta} d \cos{(\beta + \alpha')}.
    \label{eq:B_nv_position}
\end{align}
Assume we restrict rotation only around the $\hat{e}_z$ axis, so that \cite{japha2022role}:
\begin{align}
    \dot{p}_{\beta{(NV)}} & = - s \mu  \left[ \mathbf{B} \cdot \frac{\partial \hat{n}_s}{\partial \beta} + \nabla(\mathbf{B} \cdot \hat{n}_s) \cdot\frac{\partial \vec{d}}{\partial \beta}  \right],
    \label{eq:dot_p_beta_nv}
\end{align}
where $s=\pm 1$. In this case, according to diagram Fig. \ref{fig:1}(b), we have $\hat{n}_s$ and $\vec{d}$ are functions of $\beta$ (as shown in Eqs. \eqref{eq:d} and \eqref{eq:n_3}). 


If we now consider the magnetic field only in the z-axis and the NV-axis parallel to $\hat{n}_3$ ($\hat{n}_s = \hat{n}_3$), the first term of Eq.\eqref{eq:dot_p_beta_nv} becomes:
\begin{align}
    \mathbf{B} \cdot \frac{\partial \hat{n}_s}{\partial \beta} & = -{B_{nv}} \sin{\beta}\,
    \label{eq:dot_momentum_beta_1}
\end{align}
while the second term becomes:
\begin{align}
    \nabla(\mathbf{B} \cdot \hat{n}_s)
    \frac{\partial \vec{d}}{\partial \beta} & = - d \frac{\partial {B_{nv}}}{\partial z} \cos{\beta} \sin{(\beta + \alpha')}.
    \label{eq:dot_momentum_beta_2}
\end{align}
We can approximate $B_{nv} \approx B_z$, since the magnetic field gradient, together with the location of the off-centred NV, is very small $\eta d \ll B_z$. {{We set the value of $d=10$ nm according to \cite{Zhou:2024pdl} If the value of $d$ is large, there will be a mismatch in the libration angle between the two interferometer paths, causing significant rotational differences. As a result, the quantum wave packets on the two paths do not perfectly overlap, leading to a decrease in spin contrast. Moreover, to maintain high contrast, the value of $d \sin{\alpha'}$ must be very small, for example less than about 0.05 nm, which is a strict constraint on the position of the NV centre within the crystal. With a small $d$, unwanted rotation and torque can be minimized, making the interferometer more stable and the measurement results clearer.}} By combining Eq. \eqref{eq:dot_momentum_beta_1} and Eq. \eqref{eq:dot_momentum_beta_2}, we obtain the final equation of motion for the libration mode:
\begin{align}
    \Ddot{\beta }_{nv} & =  \frac{s \mu}{I} \left[ {B_{z}} \sin{\beta} - d \frac{\partial {B_z}}{\partial z} \sin{\alpha'}\right]. \label{eq:beta_nv}
\end{align}
After obtaining the complete equation for $\Ddot{\beta}$ in terms of rotation and spin-magnetic interaction, as well as the spatial equation for $\Ddot{z}$, we can now combine them to solve the resulting equation of motion for a spatial superposition. 



\subsection{$\omega_0 = 0$ case}

Based on Refs.~\cite{japha2022role,Japha:2022phw}, it is discussed that the Zeeman torque $\mu s B \cos{\beta}$ in the Hamiltonian equation indicates that the spin of the NV in the state $\ket{-1}$ results in a stable rotation when $\beta= 0$, and conversely becomes unstable when in the state $\ket{+1}$. To avoid very high $\beta$ rotation, they suggested to create a spatial superposition of the nanodiamond in the states $\ket{0}$ and $\ket{-1}$, with the spatial part of the equation of motion Eq. \eqref{eq:Ddot_z}~\footnote{Note that initial computations ignored the induced diamagnetic contribution of the nanodiamond~\cite{Japha:2022phw}; however, this was corrected in \cite{Zhou:2024pdl} and in this current analysis.}
\begin{align}
    \Ddot{z} = - \frac{s \mu \Tilde{\eta}(t)}{m} \cos{\beta}  - \frac{\chi_\rho \Tilde{\eta}^2(t)}{\mu_0} (Z_0 - z),
    \label{eq:z_no_initial_velocity}
\end{align}
\begin{figure*}
    \centering 
    \includegraphics[width=1\textwidth]{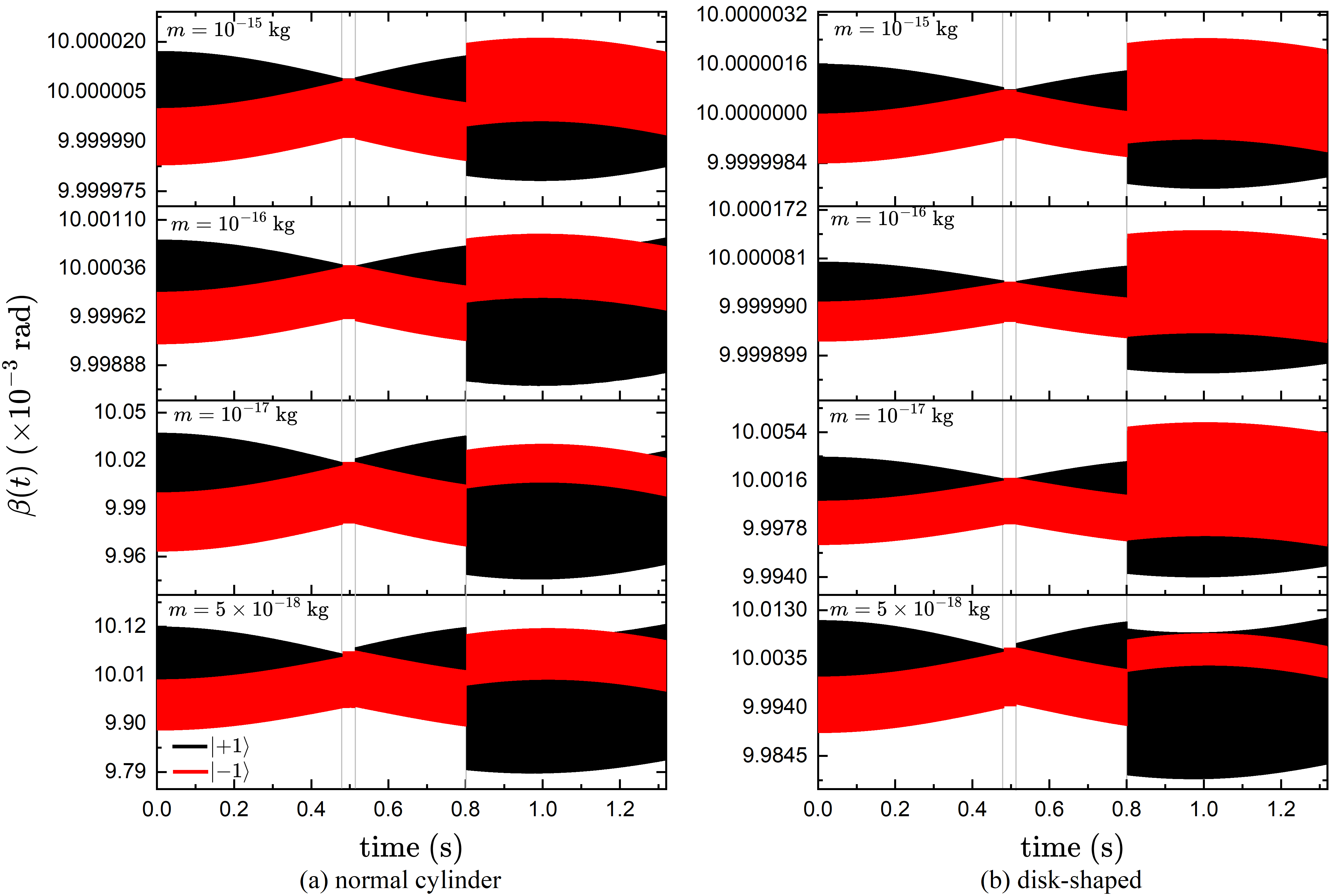}      
    \caption{{Solution of Eq. \eqref{eq:double_dot_beta_solve} with initial angle and rotation are $\beta_0 = 0.01 \ \text{rad}$ and $\omega_0 = 2\pi \times 10 \ \text{kHz}$, respectively. black and red trajectories represent libration mode $\beta(t)$ for spin-up and -down, respectively. Furthermore, we represent two forms for each set of cylindrical masses based on the diameter-to-height ratio with values $D/L = 1$ for (a) normal cylinder, and $D/L = 10$ (b) for disk-shaped (refer to the Fig. \ref{fig:1} (b) and (c)). In this plot, we set $m =  5 \times 10^{-18} - 10^{-15} \  \text{kg}$, $B_0 = 100 \ \text{G}$, $B_1 = 1 \ \text{G} $, and $\eta = 0.45G/\mu m$, for off-center distance $d = 10 \ \text{nm}$ and fix angle (NV-axis and d) $\alpha' = \pi/6$. We also set four different time stages, $\tau_1 = 0.482 \ \text{s}$ and $\tau_2 = 0.514 \ \text{s}$ are the magnetic field profile time configurations. Next the $t_{flip} = 0.8022 \ \text{s}$ is the flip NV-state time between, $\ket{+1}$ and $\ket{-1}$. And the last is $t_{closed} = 1.320 \ \text{s}$, which is the spatial trajectory of the two-stage SGI. It appears that the shape of the cylinder significantly influences the libration mode $\beta$ at small masses ($\sim10^{-18} - 10^{-17}$ kg), where the libration angle amplitude becomes very large in both spin states. Conversely, in the same shape, the amplitude stabilizes during the massive mass transition ($10^{-16} - 10^{-15}$ kg). Meanwhile, for normal-shaped cylinders and disk-shaped ones, they exhibit the same trend for the libration mode equation $\beta$. The key difference is that as the nanorotor's mass increases, the amplitude decreases, indicating that larger masses find it more challenging to perform a wobble motion.}}
    \label{fig:beta(t)}
\end{figure*}
\begin{figure*}
    \centering
    \includegraphics[width=1\linewidth]{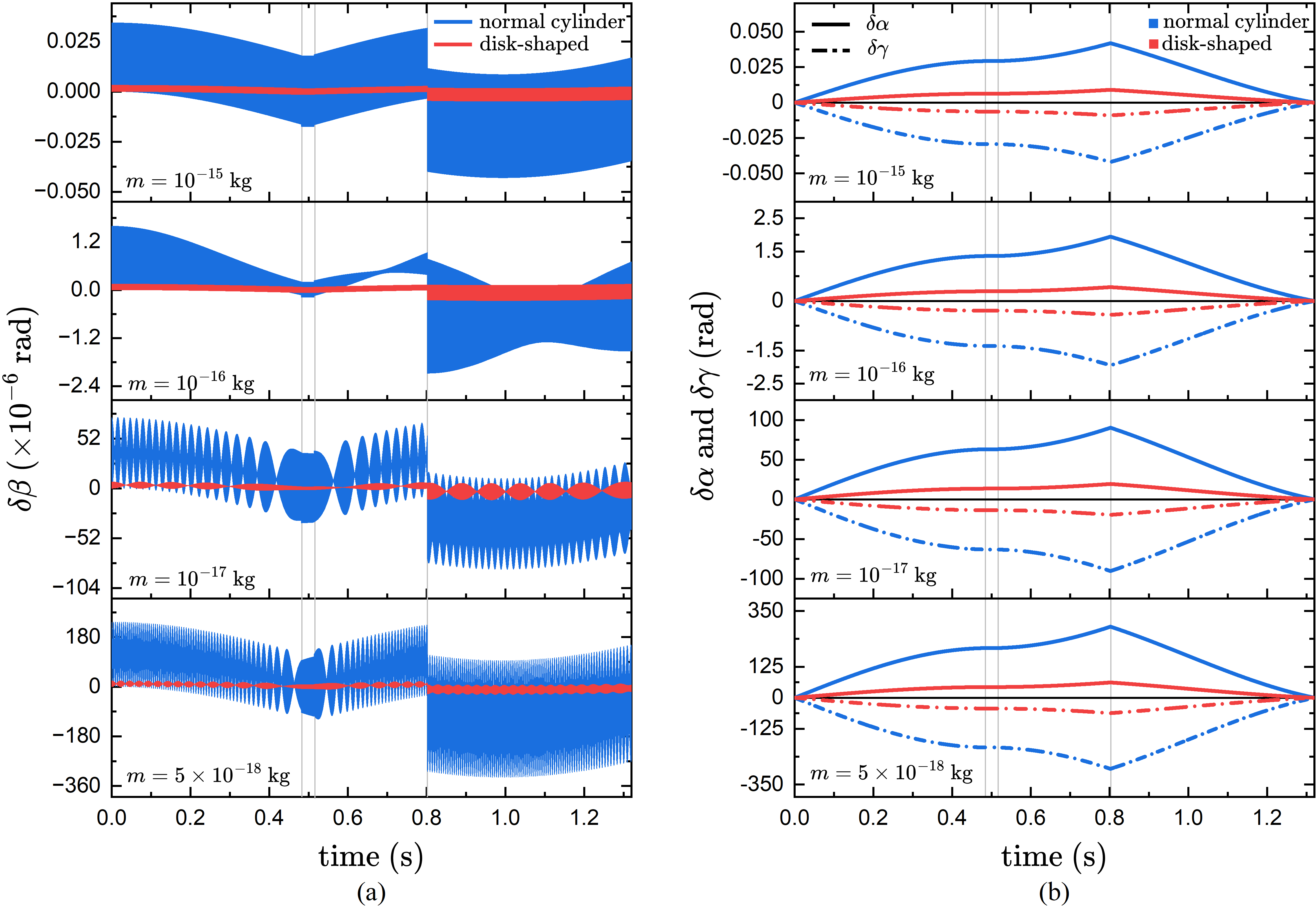}
    \caption{{(a) The angle of mismatch ($\delta\beta = \beta_{+} - \beta_{-}$) from the solution of Equation \eqref{eq:double_dot_beta_solve}. And (b) the mismatch between the angles in states $\ket{+1}$ and $\ket{-1}$ ($\delta\{\alpha,\gamma\} = \{\alpha,\gamma\}_{+} - \{\alpha,\gamma\}_{-}$) from the solution of Eq. \eqref{eq:dot_alpha_solve} and \eqref{eq:dot_gamma_solve}, with the same parameters as in Fig. \ref{fig:beta(t)}. It can be seen that the libration mode mismatch has a very large difference between the normal cylindrical shape (where $D/L= 1$) and the disk-shaped ($D/L= 10$), although they have the same mass, where the disk shape has high stability. All trends show the symmetric relationship of $\delta\alpha$ and $\delta\gamma$. The relationship $\delta\alpha \approx-\delta\gamma$ becomes a very important feature in determining the contrast of SGI. In all these examples, the embedded spin is located at a distance $d=10$ nm and angle $\alpha' = \pi/6$ from the C.O.M, with an initial angle and rotation of $\beta_0 = 0.01 \, \text{rad}$, $\omega_0 = 2\pi \times 10 \, \text{kHz}$.}}
    \label{fig:alpha_gamma}
\end{figure*}
where $Z_0=B_0/\eta$, and $\tilde \eta (t) =-\eta$ during $t< \tau_1$, $\tilde \eta (t) =0$ during $\tau_1\leq t\leq \tau_2$, and $\tilde \eta (t) =\eta$ during $t\geq \tau_2$. During the period $\tau_1\leq t\leq \tau_2$, the electronic spin can be mapped to nuclear spin to enhance the coherence of the NV spin, see~\cite{Taminiau_GM}.

For the equation of  motion of the libration mode, see  Eq.\eqref{eq:beta_nv}
\begin{align}
    \Ddot{\beta }  \approx  s \omega^2(t) \left[ {\beta} - \frac{\Tilde{\eta}(t)d \sin{\alpha'}}{B_z} \right],
    \label{eq:beta_no_initial_velocity}
\end{align}
where $\omega(t) = \sqrt{\mu B_z/I}$ as the libration frequency. The spin takes values $s = \{ 0, -1\}$ in
both Eqs.~\eqref{eq:z_no_initial_velocity}\eqref{eq:beta_no_initial_velocity}. Ref.~\cite{Japha:2022phw} showed that these are the two stable states, $s=+1$ suffers from the tachyonic instability.

The initial conditions to solve both differential equations are that the initial states are in the ground state in the trap so that $z(0)=\dot{z}(0)=\beta(0)=\dot{\beta}(0)=0$. We can find spatial trajectories for the two states, but the size of the superposition suffers compared to the $\ket{-1}$ and $\ket{+1}$ states; see the results in Appendix \ref{appendix:B}). Hence, we will concentrate on the $\omega_0\neq 0$ case, where both stable $\ket{-1}$ and $\ket{+1}$ states are prepared to create a spatial superposition.

\subsection{$\omega_0\neq 0$ }

This is the case of our interest, where we first analyse all the Euler angles $\alpha(t),~\beta(t),~\gamma(t)$. For the spatial motion, we will use Eq.\eqref{eq:z_no_initial_velocity}, with
$z(0)=\dot{z}(0)=0$ initial conditions.

Note that the complete spin part of the Hamiltonian is given by
combining $H_t=T_{rot}+H^{eff}_{NV}$, see Eq.\eqref{spin-Ham0} and \eqref{spin-Ham1}:
\begin{small}
\begin{align}
    H_t = \frac{p_\beta^2}{2I} + \frac{p_\gamma^2}{2I_3} + \frac{(p_\alpha - p_\gamma \cos{\beta})^2}{2I\sin^2{\beta}} + s\mu {B_{nv}} \cos{\beta}.
    \label{eq:H_with_NV}
\end{align}
\end{small}

Now we can re-write the equations of motion of $\beta$, as:
\begin{small}
\begin{align}
    \Ddot{\beta} & = \frac{(p_\alpha - p_\gamma \cos{\beta})(p_\alpha \cos{\beta} - p_\gamma)}{I^2 \sin^3{\beta}} + \frac{s \mu }{I} {B_{nv}}\sin{\beta}.
    \label{eq:EoM_beta}
\end{align}
\end{small}
The canonical momenta of $\alpha$ and $\gamma$ are conserved because there is no external magnetic force acting along these directions.  The initial rotation is about on the z-axis ($\dot{\alpha}(0) = 0$ and $\dot{\gamma}(0) = \omega_0$). Here we consider the small initial angle $\beta_0\neq 0$, so we can define~\footnote{We numerically analyse the case $\beta_0\neq 0$. We consider the misalignment angle $\beta_0$ small, which is possible experimentally. The embedded electronic spin in a defect, such as in the case of a NV center, can be implanted so that $\beta_0$ is a small angle, see~\cite{Japha:2022phw}.
}
\begin{align}
    p_\alpha & = I_3 \omega_0 \cos{\beta_0}, \label{eq:p_alpha}\\
    p_\gamma & = I_3\omega_0 \label{eq:p_gamma},
\end{align}
Substituting $p_\alpha$ and $p_\gamma$ in Eq. \eqref{eq:EoM_beta}, we obtain:
\begin{align}
    \Ddot{\beta} = & \ - \frac{\omega^2_0 }{\sin^3{\beta}} \frac{I_3^2}{I^2}\left(\cos{\beta_0} \nonumber - \cos{\beta}\right) \left( 1 - \cos{\beta_0} \cos{\beta} \right) \nonumber \\
    & + \frac{s \mu {B_{nv}}}{I} \sin{\beta},
    \label{eq:double_dot_beta_solve}
\end{align}
and for $\dot{\alpha}$ and $\dot{\gamma}$ given by:
\begin{align}
    \dot{\alpha} & = \frac{I_3}{I} \frac{\omega_0 \left(\cos{\beta_0} -  \cos{\beta} \right)}{\sin^2{\beta}}, \label{eq:dot_alpha_solve} \\
    \dot{\gamma} & = \frac{I_3}{I} \frac{\omega_0 \left(1 -  \cos{\beta_0}\cos{\beta}\right)}{\sin^2{\beta}}+\omega_0\left(1-\frac{I_3}{I}\right). \label{eq:dot_gamma_solve}
\end{align}
We can solve these equations by defining the constant parameters first, with the moment of inertia of the cylinder of radius R and height L, given by, $ I_1 =I_2 = I =  m(3R^2 + L^2)/12, \text{ and } I_3 = mR^2/2 $. 
{However, unlike the sphere case \cite{Zhou:2024pdl}, this scheme uses cylindrical-shaped nanodiamonds, considering that if we always maintain the value of $L$ fixed, as the mass increases, the shape of the cylinder changes to disk-shaped. To examine the relationship between the shape factor and the rotational stability of the cylinder, we introduced the ratio of the diameter of the cylinder to its height calculated based on its mass, where $D/L \gg 1$ indicates that the cylinder resembles a disk and $D/L \ll 1$ suggests that the cylinder takes the form of a long cylinder.}

\begin{figure}
    \includegraphics[width=0.9\linewidth]{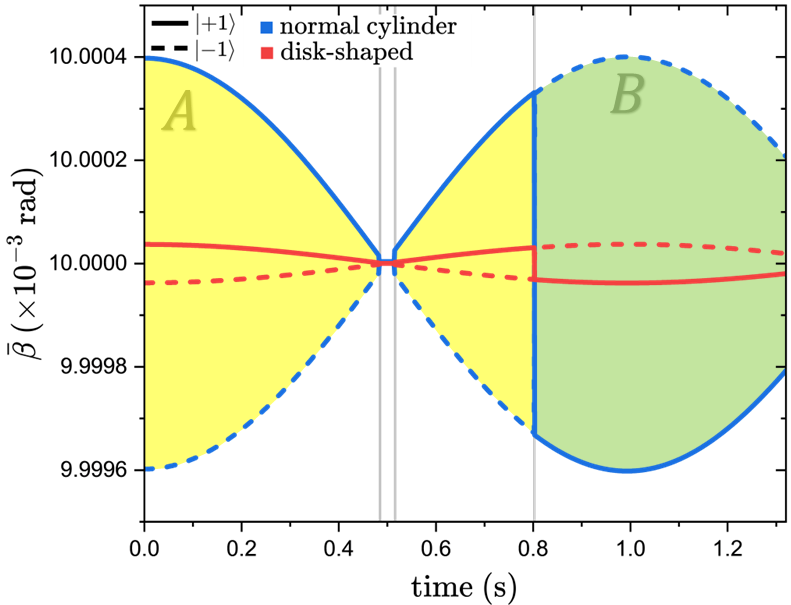}
    \caption{{These trajectories represent the two SGI arms for the equilibrium position $\bar{\beta}$ according to Eq. \eqref{eq:beta_equilibrium} of the oscillation mode $\beta(t)$. In this case, we used $m = 10^{-16} \ \text{kg}$ with two cylinder shapes: blue line for normal cylinders and red for disk-shaped, with solid and dashed lines, representing spin-up and spin-down. The area of the A region on the left side (before $t_{flip}$, mark as yellow shade, and the area B on the right side (after $t_{flip}$, mark as green shade. The shape of the normal cylinder has a very large area A and B compared to the disk even with the same mass. It is shown that the area of both sides in both modes is generally not equal. However, we can adjust the magnetic field profile parameters to ensure that $\Sigma_A \approx \Sigma_B$. In all these examples, the embedded spin is $d=10$ nm, and the angle $\alpha' = \pi/6$ from the C.O.M., with the initial angle and rotation being $\beta_0 = 0.01 \ \text{rad}$ and $\omega_0 = 2\pi \times 10 \ \text{kHz}$.}
    \label{fig:beta_equil(t)}}
\end{figure}

{Building on the analysis of the shape factor, we further determine the transition of the nanodiamond from a cylindrical shape to either a disk or a long cylinder by examining the inertia ratio $I_3/I$. Specifically, if $I_3/I \approx 2$, the nanodiamond tends to be disk-shaped, whereas if $I_3/I \ll 0$, it tends to be a long cylinder form. To accommodate this, we establish the ratio values $D/L\approx0.1$ to represent a long cylinder (see Appendix \ref{appendix:long_cylinder}), $D/L\approx1$ to represent a normal cylinder and $D/L\approx10$ for a disk-shaped.}
Using all the parameters required from~Table \ref{table:1}, we can solve differential equations \eqref{eq:Ddot_z_general} and \eqref{eq:double_dot_beta_solve} - \eqref{eq:dot_gamma_solve}   simultaneously using numerical methods with initial conditions $z(0) = \dot{z}(0) = 0$, $\beta(0) = \beta_0=0.01$ rad, $\dot{\beta}(0) = \dot{\gamma}(0) = \dot{\alpha}(0) = 0$. The numerical solutions are illustrated in Figs. (\ref{fig:beta(t)})-(\ref{fig:magnetic_field_and_z_trajectory}).


\subsection{Evolution of the libration mode $\beta$ during spatial superposition}
\label{subsection:evolution_spatial_superposition}

We perturb $\beta$ around $\beta_0$, so that $\beta - \beta_0 \ll \beta_0 \ll 1$, consequently, then Eq. \eqref{eq:double_dot_beta_solve} simplifies using Taylor series to
\begin{align}
    \Ddot{\beta} & \approx - \left( \omega^2_0 \frac{I_3^2}{I^2} - \frac{s \mu {B_z}}{I} \right) ({\beta} - \beta_0) +  \frac{s \mu {B_z}}{I} \beta_0. \label{eq:beta_minimum}
\end{align}
So we write Eq.\eqref{eq:beta_minimum} as
\begin{align}
    \Ddot{\beta} = -\omega^2(t)[\beta - \bar{\beta}(t)]. \label{eq:double_dot_beta_minimal}
\end{align}
because $\omega^2_0 \gg \mu B_z$/I (for example  $\omega^2_0\sim 10^{10} \text{Hz}$ and $\mu B_z/I \sim 10^3 \ \text{Hz}$ for nanorotor with mass $m=10^{-17} \ \text{kg}$ ). Also note that our approximation $\omega_0 <\mu B_z/I$ holds.
Therefore, we can simplify 
\begin{align}
    \omega^2(t) = \omega^2_0 \frac{I_3^2}{I^2} - \frac{s \mu {B_z(t)}}{I} \approx  \omega^2_0 \frac{I_3^2}{I^2},
    \label{cond-1-omega}
\end{align}
and the equilibrium value of $\bar \beta$ around $\beta_0$ in Eq.\eqref{eq:double_dot_beta_minimal} takes the value:
\small{
\begin{align}
    \bar{\beta}(t) = \beta_0  + \frac{s \mu {B_z}}{I\omega^2_0 \frac{I_3^2}{I^2} - s \mu {B_z}} \beta_0 \approx \beta_0  + \frac{I^2}{I_3^2} \frac{s \mu {B_z(t)}}{I\omega^2_0} \beta_0.
    \label{eq:beta_equilibrium}
\end{align}
}
Since, we are interested in a small wobble in the libration angle $\beta$, we must ensure that $\omega_0$ is such that the evolution of $\beta$ remains small over the course of the change in the magnetic field gradients in the Stern-Gerlach setup. Hence, we assume the value of $\omega_0$ such that
it satisfies all the requirements mentioned above\footnote{{Based on experiments reported by \cite{jin2024quantum, Wood2017, Wood2018}, levitated nanodiamonds can be spun up to the MHz frequency scale.}}, see Eqs.\eqref{eq:beta_minimum},\eqref{cond-1-omega}, and \eqref{eq:EdH_constrain}. Thus,
\begin{equation}
\sqrt{\frac{\mu B_z}{I}}\ll \omega_0\ll \frac{\mu B_z}{\hbar}\, ,\  {\rm and}\ \omega_0\ll \frac{\mathcal{D}\pm\mu B_z}{\hbar}.
\end{equation}
The solution for Eq.\eqref{eq:double_dot_beta_minimal} can be solved by an ansatz:
\begin{align}
    \beta(t) = A_\beta \cos{\omega t} + \bar{\beta},    
\end{align}
where initial condition $\beta(0) = \beta_0$ and $\dot{\beta}(0)=0$. We can also get the initial amplitude of the libration angle, as:
\begin{align}
    A_\beta(t) = \frac{I^2}{I_3^2} \frac{s \mu {B_z(t)}}{I\omega^2_0} \beta_0, \ \ \ & 0\leq t < t_{flip}
    \label{eq:initial_amplitude_libration_mode}
\end{align}
when $t\rightarrow t_{flip}$, we flip the NV spin between $s = +1$ and $s = -1$, so the equilibrium position of  $\bar{\beta}$ is changed, hence the change in the libration mismatch angle $\delta \bar \beta$ is given by: 
\begin{align}
    \delta\bar{\beta}(t_{flip}) & = \bar{\beta}(t_{flip},s=+1) - \bar{\beta}(t_{flip},s=-1) \nonumber \\
    & = \frac{I^2}{I_3^2} \frac{2\mu {B_z(t_{flip})}}{I\omega^2_0} \beta_0
\end{align}
We can also roughly estimate the amplitude after flipping the spin  ($t > t_{flip}$), which is now bounded by the two  amplitudes: 
\begin{align}
    |A_\beta(0)-\delta\bar{\beta}(t_{flip})| \leq A_\beta(t) \leq |A_\beta(0)+\delta\bar{\beta}(t_{flip})|
\end{align}
In our example, we can use the magnetic field profile to estimate the upper and lower bounds of $A_\beta (t> t_{flip})$.
Since our magnetic field profile satisfies $B_z(t_{closed}) \sim B_z(t_{flip}) \lesssim B_z(0) =B_0$, see Eq.\eqref{eq:magnetic_field_profile}. Hence, substituting $B_z(t_{flip})\approx B_0$, we obtain:
\begin{widetext}
\begin{align}
    \abs{A_\beta(0)-\frac{I^2}{I_3^2} \frac{2\mu {B_0}}{I\omega^2_0} \beta_0} & \lesssim  A_\beta(t>t_{flip}) \lesssim \abs{A_\beta(0)+\frac{I^2}{I_3^2} \frac{2\mu {B_0}}{I\omega^2_0} \beta_0} \nonumber \\
    \abs{A_\beta(0)-2A_\beta(0)} & \lesssim  A_\beta(t > t_{flip}) \lesssim \abs{A_\beta(0)+2A_\beta(0)} \nonumber \\
    A_\beta(0) & \lesssim A_\beta(t > t_{flip}) \lesssim 3A_\beta(0). \label{eq:amplitude_t_flip}
\end{align}
\end{widetext}
Finally, when $t=t_{closed}$, the mismatch in  $\beta$ between the two paths, e.g. $|+\rangle$ and $|-\rangle$ can be approximately determine by: 
\begin{align*}
    \delta\bar{\beta}(t_{closed}) \approx \frac{I^2}{I_3^2} \frac{2\mu {B_z(t_{closed})}}{I\omega^2_0} \beta_0\,.
\end{align*}
Base on our approximation of Eq.\eqref{eq:amplitude_t_flip}, where $ A_\beta (t_{closed}) \lesssim 3A_\beta(0) $), we can estimate the upper bound on the libration mode's oscillation amplitude, which is given by:
\begin{align}
    A_\beta (t_{closed}) \lesssim \frac{I^2}{I_3^2} \frac{3\mu {B_0}}{I\omega^2_0} \beta_0,
    \label{eq:Amplitude_t_closed}
\end{align}
Now we can finally have an estimation of the mismatch $\delta\beta$ upon closure of the two spin trajectories
with the help of $\delta\bar{\beta}(t_{closed})$ and $A_\beta (t_{closed})$, as
\begin{align}
    \delta \beta & < A_{\beta{(+1)}}(t_{closed}) + A_{\beta (-1)}(t_{closed}) + \delta\bar{\beta}(t_{closed}) \nonumber \\ 
    &\lesssim 3A_\beta(0) + 3A_\beta(0)+\frac{I^2}{I_3^2} \frac{2\mu {B_0}}{I\omega^2_0} \beta_0 \nonumber \\
    \delta\beta&\lesssim \frac{I^2}{I_3^2} \frac{8\mu B_0\beta_0}{I\omega_0^2}.
\end{align}
where $A_{\beta{(+1)}}$ and $A_{\beta{(-1)}}$ belong to the amplitude of the $\beta$ oscillations. Note that $\delta \beta$ only depends on the moment of inertia, magnetic field, initial rotational frequency, and the initial value of the libration mode. Hence, we can see that the mismatch in the libration mode is suppressed by the moment of inertia and $\omega_0^2$. This combination provides the solution to the Humpty-Dumpty problem in one-dimensional one-loop interferometry for the libration mode. This intuition will be evidenced in our numerical solution of $\beta (t)$ and $\delta \beta(t)$, see Figs.~(\ref{fig:beta(t)} \ \text{and} \ \ref{fig:alpha_gamma}(a)).

In Figs.~(\ref{fig:beta(t)}), we show the evolution of $\beta(t)$ for states $|+1\rangle \ \text{and}  ~|-1\rangle$ for different masses ranging from $5 \times10^{-18} {~\rm kg}\leq m\leq 10^{-15}{~\rm kg}$ and present this in two different shapes, which are normal cylinder and disk-shapes. There are various jumps in the evolution marked by faint vertical lines. These jumps are due to the change in the magnetic field profile. Initially, we have an acceleration of the nanorotor as a result of a positive magnetic field gradient. We have a free motion of $\beta$, and then deceleration of the nanorotor's motion, which affects the evolution of $\beta$ due to a negative magnetic field gradient (required for the recombination of the $|-\rangle$ and $|+\rangle$ arms of the trajectories). The oscillations are rapid, with a frequency that matches close to the initial imparted frequency on the nanorotor, e.g. $\omega_0\sim 2\pi \times 10$KHz.  For illustration, we have taken $B_0=100$ G, $B_1=1$ G and $\eta=0.45 {\rm G/\mu m}$ in our numerical simulations, and the time stages are $\tau_1=0.482,~\tau_2=0.514,~\tau_{flip}=0.8022$ s, and we close the interferometer at $t=1.320$ s. In the panel Fig.(\ref{fig:alpha_gamma} (a)), we consider the mismatch in the libration angle $\delta \beta$ for the states $|-1\rangle$ and $|+1\rangle$ for masses ranging $5\times 10^{-18} {~\rm kg}\leq m\leq 10^{-15}{~\rm kg}$, with blue lines for the normal cylinder and red for the disk-shaped.
\subsection{Evolution of $\alpha$ and $\gamma$ during spatial superposition}
\begin{figure*}
    \centering
    \includegraphics[width=1\linewidth]{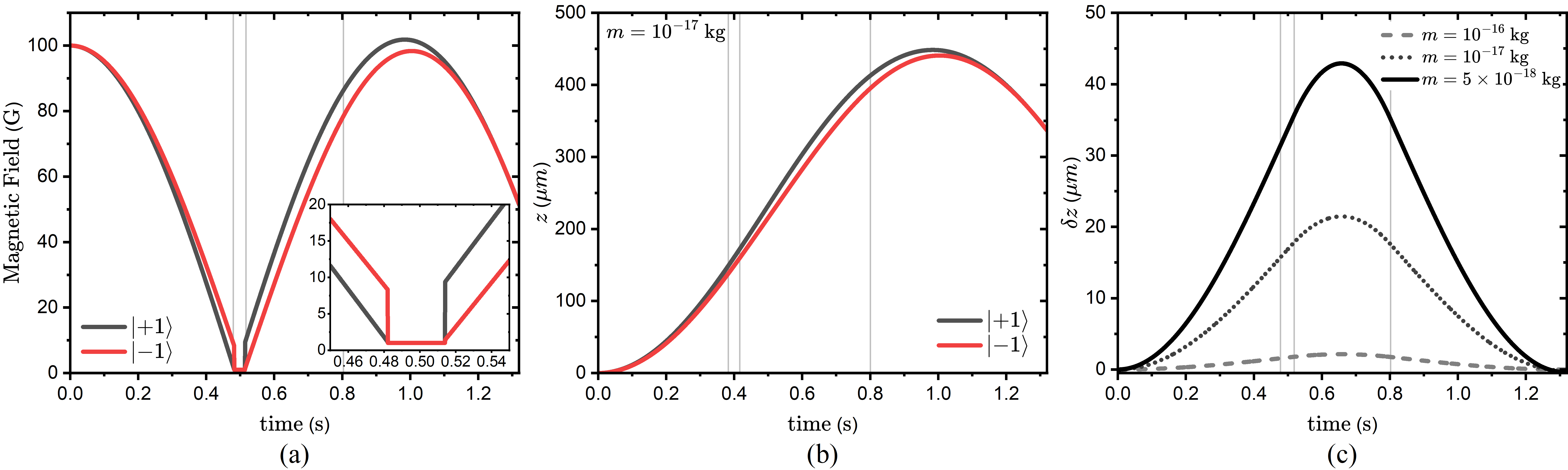}
    \caption{{
    (a) The magnetic field profile that we used for each arm of the interferometer as a function of time has constant parameters $B_0$, $B_1$, and $\eta$, set at values of $10^{-2} T$, $10^{-4} \ \text{T}$, and $0.45 \times 10^2 \ \text{T/m}$ respectively, for the time conditions $\tau_1 = 0.482 \ \text{s}$ and $\tau_2 = 0.514 \ \text{s}$, as referred to in Ref. \cite{Marshman2022}. (b) The spatial trajectories of $z(t)$ for both spin-up (black) and -down(red) for mass with mass $m=10^{-17} \ \text{kg}$ and using magnetic field profile from Eq. \eqref{eq:magnetic_field_profile}, and (c) represents superposition size between two arms varies with mass $m = 5\times 10^{-18} - 10^{-16} \ \text{kg}$ according to the solution of EoM Eq. \eqref{eq:Ddot_z} with neglected value of E. The value of $\mathcal{D}=2.87 h \times 10^9$ Hz,  and $\chi_\rho = -6.2\times 10^{-9} \text{m}^3/\text{kg}$. The maximum value of superposition is approximately $ 43 \ \mu m$ for an extremely light nanorotor with a mass of $m = 5\times 10^{-18}$ kg and decreases to approximately $2.1 \ \mu m$ for a heavier mass of $m = 10^{-16}$ kg. As we see, the trajectory of two arms has a good approximation of the superposition. In all these examples the embedded spin is located at a distance $d=10$ nm and angle $\alpha' = \pi/6$ from the C.O.M, with initial angle and rotation are $\beta_0 = 0.01 \ \text{rad}$, $\omega_0 = 2\pi \times 10 \ \text{kHz}$. Note that, for the z-direction trajectory, there is no significant difference in superposition size between a normal cylinder and a disk-shaped one. We set $L= 100$ nm, and the range radius of the cylindrical nanorotor is $67.4 - 301.6$ nm as the mass increase from $5\times 10^{-18} - 10^{-16}$ kg, indicating transition of cylinder shape from normal cylinder to disk-shaped. 
    }}
    \label{fig:magnetic_field_and_z_trajectory}
\end{figure*}
Similarly to Eq.~\eqref{eq:beta_minimum}, we consider the equations of motion for the precession angle $\alpha$ and the rotation angle $\gamma$. Thankfully, for these two variables, there are no external forces, hence the solutions simplify a lot, and also Eqs.\eqref{eq:dot_alpha_solve} and \eqref{eq:dot_gamma_solve}: 
\begin{align}
    \dot{\alpha} & \approx \frac{I_3}{I} \frac{\omega_0}{\beta_0} (\beta - \beta_0) \label{eq:alpha_minimum}  \\
    \dot{\gamma} & \approx -\frac{I_3}{I} \frac{\omega_0}{\beta_0} (\beta - \beta_0)+\omega_0\left(1-\frac{I_3}{I} \right) \label{eq:gamma_minimum}.
\end{align}

Note that $\dot \alpha=-\dot \gamma+\omega_0(1-I_3/I)$ and their respective evolution depend on $\beta$, which we now know. Therefore, we can also evaluate the mismatch for $\alpha(t)$ at any given moment of time, which is given by $\delta \alpha (t)=-\delta \gamma (t)$, and 
\begin{align}
    \delta \alpha(t) &\approx \frac{I_3}{I} \frac{\omega_0}{\beta_0} \int_0^{t_{closed}} dt [\bar{\beta}_{+}(t) - \bar{\beta}_{-}(t)]  \nonumber \\
    &\approx \frac{I_3}{I} \frac{\omega_0}{\beta_0} (\Sigma_A-\Sigma_B),
    \label{eq:mismatch_delta_alpha}
\end{align}
{where the $\bar{\beta}_+$ and $\bar{\beta}_-$ represent blue and red curves in Fig. \ref{fig:beta_equil(t)}. In the same figure, we define $\Sigma_A=\Sigma(t_{flip})$ as representing the area of region A (yellow) on the left side (before $t_{flip}$) and $\Sigma_B(t_{closed}-t_{flip})$ is region B (green) on the right side (after $t_{flip}$). Note that we have replaced the integration by the area covered by the left and right parts of the trajectories since $\beta$ oscillating so rapidly over the time scale of the closure of the interferometry that we can take the amplitude of the respective $\beta$'s evaluating the integration, which yields the area covered by the left and right paths in the integration. We also see from Fig.~\ref{fig:alpha_gamma}(b) that the mismatch of the precession and the rotation angles has a relation $\delta\alpha = -\delta \gamma$. Note that since both $\delta \alpha$ and $\delta \gamma$ are proportional to the moment of inertia $I_3$, therefore a smaller $I_3$ will yield a smaller change in $\alpha$ and $\gamma$.}

\subsection{Spatial superposition including rotation}

It remains to be seen how the evolution of the superposition evolves with time for different masses including the rotation of the nanorotor in our case. 
We perform the spatial evolution of the equation of motion for the $z$ degree of freedom coupled to the rotation of the nanoparticle by following Eq.~\eqref{eq:Ddot_z}. This part of the computation we perform numerically. Our results are shown in Fig.~\ref{fig:magnetic_field_and_z_trajectory}.

After demonstrating that the geometric configuration of the cylinder minimises the mismatch in the libration mode $\beta$, we now investigate how effectively this configuration can generate spatial superposition. In panel Fig.~\ref{fig:magnetic_field_and_z_trajectory} (a), we show the magnetic field profile (in Gauss) over time. We show the change in the magnetic field for both $|+1\rangle $ and $|-1\rangle$ states. In the initial stages, the nanorotor's C.O.M experiences different forces for the $|-1\rangle$ and $|+1\rangle$ states; hence, the difference in the potential energy and the force yields a slight spatial shift in the two states to create the superposition.
During the time when there is no magnetic field gradient, the particle's quantum evolution is that of a free evolution. In Fig.~\ref{fig:magnetic_field_and_z_trajectory}(b), we show how the paths for the $|-1\rangle$ state and the $|+1\rangle $ state evolve over time. Here we ensure that for the parameters, both the position and the momentum overlap exactly at time $t=1.32$ s. Panel (b) shows this evolution for mass $m=10^{-17}$ kg, for which we create the maximum superposition size $\delta z\sim 40 {\rm \mu m}$.  Of course, for numerical sake, we consider the specifications of those of the nanodiamond, but our analysis is very generic and will hold for any diamagnetic cylindrical nanorotor.  

{In panel Fig.~\ref{fig:magnetic_field_and_z_trajectory} (c), we show the superpositions by varying the mass. As we increase the mass, the size of the superposition decreases dramatically. We set the cylinder height at $L = {100}$ nm. As the mass increases from ${5\times10^{-18}}$ kg to ${10^{-16}}$ kg, the radius increases from 67.4 nm to 301.6 nm, causing the cylinder's shape to transition from normal to disk-shaped. However, because the difference in the libration mode $\beta$ is just $10^{-1}$ rad (see Figs. \ref{fig:beta(t)} and \ref{fig:beta_equil(t)}) for the trajectory in the z direction, the superposition size does not differ significantly between a normal cylinder and a disk-shaped one.}

\section{Spin Contrast in one-loop interferometry}
\label{chapter:spin_contrast}

To determine the interferometric contrast, for example, the spin contrast, as we read the final outcome through the spin population, we need to assess the Humpty-Dumpty problem \cite{Englert,Schwinger,Scully}. This was also analysed in Refs.~\cite{Japha:2022phw,Zhou:2024pdl}. Here we will provide the details of the computation and apply that to our cylindrical nanorotor (for both the normal cylinder and disk-shaped).

We consider the canonical quantisation procedure, which is consistent with Ref.\cite{Barut1992}. It is well known that the angular momentum part of the commutation relationships is given by: $[\hat{\beta},\hat{p}_\beta]= [\hat{\alpha},\hat{p}_\alpha] = [\hat{\gamma},\hat{p}_\gamma] = i\hbar \ \text{and} \ [\hat{p}_\beta,\hat{p}_\alpha] = [\hat{p}_\beta,\hat{p}_\gamma] = [\hat{p}_\alpha,\hat{p}_\gamma] = 0$.

We examine the rotational Hamiltonian again for the spin contrast given in Eq.~\eqref{eq:H_with_NV}. We briefly recall it here: 
\begin{align}
    H_t = \frac{p_\beta^2}{2I} + \frac{p_\gamma^2}{2I_3} + \frac{(p_\alpha - p_\gamma \cos{\beta})^2}{2I\sin^2{\beta}} + s\mu {B_{nv}} \cos{\beta} \nonumber,
    \label{spin-Ham-HD}
\end{align}
After determining that $\beta(t)$ has an equilibrium value \(\bar{\beta}\), see Eq.\eqref{eq:beta_equilibrium}, we also introduced the momenta \(p_\alpha\) and \(p_\gamma\) and their equations of motion in Eqs.~\eqref{eq:p_alpha} and \eqref{eq:p_gamma}. To understand the Humpty-Dumpty problem, we will need to perturb the interaction Hamiltonian and study the fluctuations around the equilibrium state for the libration mode~\footnote{In sections \ref{chapter:spin_contrast} and \ref{chapter:spin_contrast_with_temperature}. 
we followed and modified the derivation of the spin contrast for the cylinder case. See Appendix D in ref. \cite{Zhou:2024pdl}, where the authors discussed the spherical geometry. }. We first shift the operators involved in  Eq.\eqref{spin-Ham-HD}:
\begin{align}
    \hat{\beta}' & = \hat{\beta} - \bar{\beta}, \\
    \hat{p}'_\alpha & = \hat{p}_\alpha - \langle{\hat{p}_\alpha}\rangle \\
    \hat{p}'_\gamma & = \hat{p}_\gamma - \langle{\hat{p}_\gamma}\rangle
\end{align}
where $\beta$ at the equilibrium position is defined in Eq.~\eqref{eq:beta_equilibrium} as $\bar{\beta} = \beta_0 + (I/I_3)^2 s\mu B_z/(I\omega_0^2)$. The expectation values of the conjugate momentum for the angles $(\alpha, \gamma)$ are given by $\langle{\hat{p}_\alpha}\rangle = \langle{\hat{p}_\gamma}\rangle \cos{\beta_0}$ and $\langle{\hat{p}_\gamma}\rangle = I_3 \omega_0$. Now we can express the Hamiltonian Eq.\eqref{eq:H_with_NV} for the rotational dynamics as a function of these shifted operators, as follows:
\begin{align}
    H_t \approx & \ \frac{p_\beta^2}{2I} + \frac{I_3^2}{I^2} \frac{I\omega^2_0}{2} \beta'^2 - f(p'_\alpha,p'_\gamma)\beta' + g(p'_\alpha,p'_\gamma,t)
    \label{eq:hamiltonian_with_shifted_operator}
\end{align}
where the functions $f$ and $g$ are given by:
\begin{small}
    \begin{align}
        f(p'_\alpha,p'_\gamma) = & \ \frac{(p'_\alpha - p'_\gamma)^2}{I\beta_0^3} + \frac{I_3\omega_0}{I\beta_0}(p'_\alpha - p'_\gamma), \label{eq:H_F} \\
        g(p'_\alpha,p'_\gamma,t) = & \ \frac{(p'_\alpha - p'_\gamma)^2}{2I\beta_0^2} - \omega_0p'_\gamma  - \frac{I}{I_3} \frac{s\mu B_z(t)}{I\omega_0}(p'_\alpha - p'_\gamma). \label{eq:H_G(t)}
\end{align}
\end{small}
Given that $\langle \beta' \rangle \ll \beta_0 \ll 1$, we can neglect the higher-order contributions of $\beta'$. Also, we have $\Delta p_\alpha \approx \Delta p_\gamma \approx \hbar \ll I\omega_0$. Hence, we assume that the momentum fluctuations are small compared to the inertial rotation in our analysis. We further assume that the quantum wave packets of $\alpha$ and $\gamma$ in the momentum space are Gaussian wave packets with deviations $\Delta p_\alpha$ and $\Delta p_\gamma$, respectively. 
The quantum state of the libration mode $\beta$ is set as the ground state $\ket{0}_{\bar{\beta}_0}$ in a harmonic trap with frequency $\omega_0$ and equilibrium position $\bar{\beta}_0$. Thus, we can express the wave function of the angular degree of freedom as follows:
\begin{align}
    \ket{\Psi(0)} = & \ \int \frac{dp'_\alpha dp'_\gamma}{\sqrt{2\pi \Delta p_\alpha \Delta p_\gamma}} e^{-\frac{p'^2_\alpha}{4\Delta p^2_\alpha}} e^{-\frac{p'^2_\gamma}{4\Delta p^2_\gamma}} \ket{p'_\alpha,p'_\gamma} \otimes \ket{0}_{\bar{\beta}_0}.
\end{align}
With our shifted spin Hamiltonian, Eq. \eqref{eq:hamiltonian_with_shifted_operator}, we can rewrite the initial equilibrium position $\bar{\beta}(0)$ for the libration mode, $\beta$, as
\begin{align}
    \bar{\beta}_0(p'_\alpha,p'_\gamma) = \beta_0 + \frac{f(p'_\alpha,p'_\gamma)}{I\omega_0^2}
    \label{eq:initial_equilibrium_position}
\end{align}
The equilibrium position of the harmonic trap is influenced by the quantum fluctuations in $p'_\alpha$ and $p'_\gamma$.

At the start of the interferometer, the spin state is initialised as a superposition of $s = +1$ and $s = -1$. Consequently, $\bar{\beta}$ shifts from the initial equilibrium position $\bar{\beta}_0$, causing the ground state $\ket{0}_{\bar{\beta}_0}$ to become a coherent state in the harmonic trap with a new equilibrium position; see~\cite{Pedernales:2020nmf}. The new equilibrium position is given as follows:
\begin{small}
\begin{align}
    \bar{\beta}(s,t,p'_\alpha,p'_\gamma) = \beta_0 + \frac{I^2}{I_3^2}\frac{s\mu B_z(t)\beta_0}{I\omega_0^2}+\frac{f(p'_\alpha,p'_\gamma)}{I\omega_0^2}
    \label{eq:new_equilibrium_position}
\end{align}
\end{small}
Thus, the quantum evolution of the wave packet can be described by $\ket{\Psi(t)} = \exp\left[\left(-{i}/{\hbar} \right)\int H_t dt \right] \ket{\Psi(0)}$, and substituting the definition of $H_t$ from Eq. \eqref{eq:hamiltonian_with_shifted_operator}, we obtain the following expression:
\begin{widetext}
\begin{align}
    \ket{\Psi(t)} = \ \int \frac{dp'_\alpha dp'_\gamma}{\sqrt{2\pi \Delta p_\alpha \Delta p_\gamma}} e^{-\frac{p'^2_\alpha}{4\Delta p^2_\alpha}} e^{-\frac{p'^2_\gamma}{4\Delta p^2_\gamma}} \times \exp \left[ -\frac{i}{\hbar} \int dt g(p'_\alpha,p'_\gamma,t) \right] \ket{p'_\alpha,p'_\gamma} \otimes \ket{\kappa(t)}_{\bar{\beta}}
    \label{eq:wave_packet_of quantum_evolution}
\end{align}    
\end{widetext}
The last term of eq. \eqref{eq:wave_packet_of quantum_evolution}, $\ket{\kappa(t)}_{\bar{\beta}}$, represents the coherent state; as we evolve in time, the coherent states will become
\begin{widetext}
\begin{align}
    \ket{\kappa(t)}_{\bar{\beta}}  = \exp\left[-\frac{i}{\hbar} \int dt \left( \frac{p_\beta^2}{2I} + \frac{I_3^2}{I^2} \frac{I \omega_0^2}{2} \beta'^2 - f(p'_\alpha,p'_\gamma)\beta' \right) \right] \ket{0}_{\bar{\beta}_{0}}
     = \ket{- \sqrt{\frac{I \omega_0}{2\hbar}} \frac{I^2}{I_3^2} \frac{s\mu B_z(t) \beta_0}{I\omega_0^2}e^{-i\omega_0 t}}_{\bar{\beta}}.
    \label{eq:coherent_state_def}
\end{align}
\end{widetext}
Note that from Eqs.~\eqref{eq:initial_equilibrium_position} and \eqref{eq:new_equilibrium_position}, we see that both $\bar{\beta}(s,t,p'_\alpha,p'_\gamma)$ and $\bar{\beta}_0(p'_\alpha,p'_\gamma)$ are shifted by $f(p'_\alpha,p'_\gamma)/(I\omega_0^2)$. However, since we assume that the quantum fluctuations of momentum $p_\alpha$ and $p_\gamma$ are small compared to the classical rotation ($\Delta p_\alpha \approx \Delta p_\gamma \approx \hbar \ll I\omega_0$), hence, the initial amplitude of the libration mode, $A_\beta(0)$ (in Eq. \eqref{eq:initial_amplitude_libration_mode}), was not affected by quantum fluctuations in momentum $p_\alpha$ and $p_\gamma$. From the 
Eq. \eqref{eq:coherent_state_def}, we find the initial amplitude of $\abs{\kappa(0)} $:
\begin{align}
    \abs{\kappa(0)} = \sqrt{\frac{I \omega_0}{2\hbar}} \frac{I^2}{I_3^2} \frac{\mu B_0 \beta_0}{I\omega_0^2} =  \sqrt{\frac{I \omega_0}{2\hbar}}A_\beta(0),
\end{align}
The evolution of the amplitude
$\abs{\kappa}$ of the coherent state can be determined through a classical amplitude estimation; see Section \eqref{subsection:evolution_spatial_superposition}. Let us recall the upper bound of the final amplitude in Eq. \eqref{eq:Amplitude_t_closed}. Therefore, we have the final amplitude of $\abs{\kappa}$ becomes:
\begin{align}
    \abs{\kappa(t_{closed})} & < \sqrt{\frac{I \omega_0}{2\hbar}} [3A_\beta(0)] \nonumber \\
    & <  \sqrt{\frac{I\omega_0}{2\hbar}} \frac{I^2}{I_3^2} \frac{3\mu B_0 \beta_0}{I\omega_0^2}
    \label{eq:coherence_at_t_closed}
\end{align}
%
Note that since the Hamiltonian of the quantum state for $\alpha$ and $\gamma$ consists only of the quadratic terms of $\hat{p}'_\alpha$ and $\hat{p}'_\gamma$, the wave packets of $\alpha$ and $\gamma$ remain in Gaussian form during the evolution of the harmonic oscillator. 
Since $[\hat{p}'_\alpha,\hat{H}] = [\hat{p}'_\gamma,\hat{H}] = 0 $, the momentum uncertainties $\Delta p_\alpha$ and $\Delta p_\gamma$ remain invariant.

We can now calculate the spin contrast for the rotational dynamics based on the solution of the angular quantum state $\ket{\psi(t)}$, as follows:
\begin{widetext}
\begin{align}
    C = &  \ \abs{\braket{\Psi_-(t_{closed})}{\Psi_+(t_{closed})}} \nonumber \\
    = & \ \int\frac{dp'_\alpha dp'_\gamma dp''_\alpha dp''_\gamma}{2\pi\Delta p_\alpha \Delta p_\gamma} \times e^{-\frac{(p'^2_\alpha)+(p''^2_\alpha)}{4\Delta p^2_\alpha}} e^{-\frac{(p'^2_\gamma)+(p''^2_\gamma)}{4\Delta p^2_\gamma}}  \exp{-\frac{i}{\hbar} \int_0 ^{t_{closed}} dt \left[ g_-(p'_\alpha,p'_\gamma,t)-g_+(p''_\alpha,p''_\gamma,t)\right]} 
     \nonumber \\
    & \times \braket{p'_\alpha,p'_\gamma}{p''_\alpha,p''_\gamma}  \abs{\prescript{}{\bar{\beta}_{-}} {\braket{\kappa_- (p'_\alpha,p'_\gamma,t_{closed})}{\kappa_+ (p'_\alpha,p'_\gamma,t_{closed})}}_{\bar{\beta}_{+}}} \nonumber \\
    = & \ \int\frac{dp'_\alpha dp'_\gamma}{2\pi\Delta p_\alpha \Delta p_\gamma} \times e^{-\frac{p'^2_\alpha}{2\Delta p^2_\alpha}} \times e^{-\frac{p'^2_\gamma}{2\Delta p^2_\gamma}} \exp{-\frac{i}{\hbar} \int_0 ^{t_{closed}} dt \left[ g_-(p'_\alpha,p'_\gamma,t)-g_+(p'_\alpha,p'_\gamma,t)\right]} 
    \nonumber \\
    & \times \abs{\prescript{}{\bar{\beta}_{-}} {\braket{\kappa_- (p'_\alpha,p'_\gamma,t_{closed})}{\kappa_+ (p'_\alpha,p'_\gamma,t_{closed})}}_{\bar{\beta}_{+}}}
    \label{eq:Contrast_def}
\end{align}
\end{widetext}
For the exponential part of Eq.\eqref{eq:Contrast_def}, we can substitute Eq.\eqref{eq:H_G(t)}, and write down the solution as a term of the mismatch between the classical trajectories of $\alpha$ between the left and right part of the SGI, $\delta \alpha$ in Eq. \eqref{eq:mismatch_delta_alpha}, becomes:
\begin{align}
    & \int_0 ^{t_{closed}} dt \left[ g_-(p'_\alpha,p'_\gamma,t)-g_+(p'_\alpha,p'_\gamma,t)\right] \nonumber\\
    & = \int_0 ^{t_{closed}} dt \left[ \frac{I}{I_3} \frac{\mu s B_z}{I\omega_0} \bigg|_- -  \frac{I}{I_3} \frac{\mu s B_z}{I\omega_0} \bigg|_+ \right](p'_\alpha - p'_\gamma) \nonumber\\
    & = \left\{ \frac{I_3}{I} \frac{\omega_0}{\beta_0} \int_0 ^{t_{closed}} dt \left[ \bar{\beta}_- -  \bar{\beta}_+ \right] \right\}(p'_\alpha - p'_\gamma) \nonumber\\
    & = \delta \alpha \times (p'_\alpha - p'_\gamma). \label{eq:integral_G_term}
\end{align}
Because of relation $\delta \alpha \approx - \delta \gamma$, we can write down the integral result of Eq. \eqref{eq:integral_G_term} as
\begin{small}
\begin{align}
    \int_0 ^{t_{closed}} dt \left[ g_-(p'_\alpha,p'_\gamma,t)-g_+(p'_\alpha,p'_\gamma,t)\right] = \delta \alpha p'_\alpha + \delta\gamma p'_\gamma.
\end{align}
\end{small}
Next, for the term of the inner product $\abs{\prescript{}{\bar{\beta}_{-}} {\braket{\kappa_- (p'_\alpha,p'_\gamma,t_{closed})}{\kappa_+ (p'_\alpha,p'_\gamma,t_{closed})}}_{\bar{\beta}_{+}}}$ of two coherent states is determined by the distance between the complex variables $\kappa_-$ and $\kappa_+$, written as ~\cite{Zhou:2024pdl}:
\begin{align}
    \abs{_{\bar{\beta}_{L}} \braket{\kappa_- 
    }{\kappa_- 
    }_{\bar{\beta}_{+}}}
    = \exp{-\frac{1}{2}\abs{\kappa_- - (\kappa_+ +\delta X)}^2},
    \label{eq:coherence_state_normalization}
\end{align}
where
\begin{align}
    \delta X & = \sqrt{\frac{I\omega_0}{2\hbar}}\abs{\bar{\beta}_- - \bar{\beta}_+} = \sqrt{\frac{I\omega_0}{2\hbar}} \frac{I^2}{{I_3}^2} \frac{2\mu B_z (t_{closed})\beta_0}{I {\omega_0}^2}.
    \label{eq:delta_X}
\end{align}

\begin{figure*}
    \centering
    \includegraphics[width=1\linewidth]{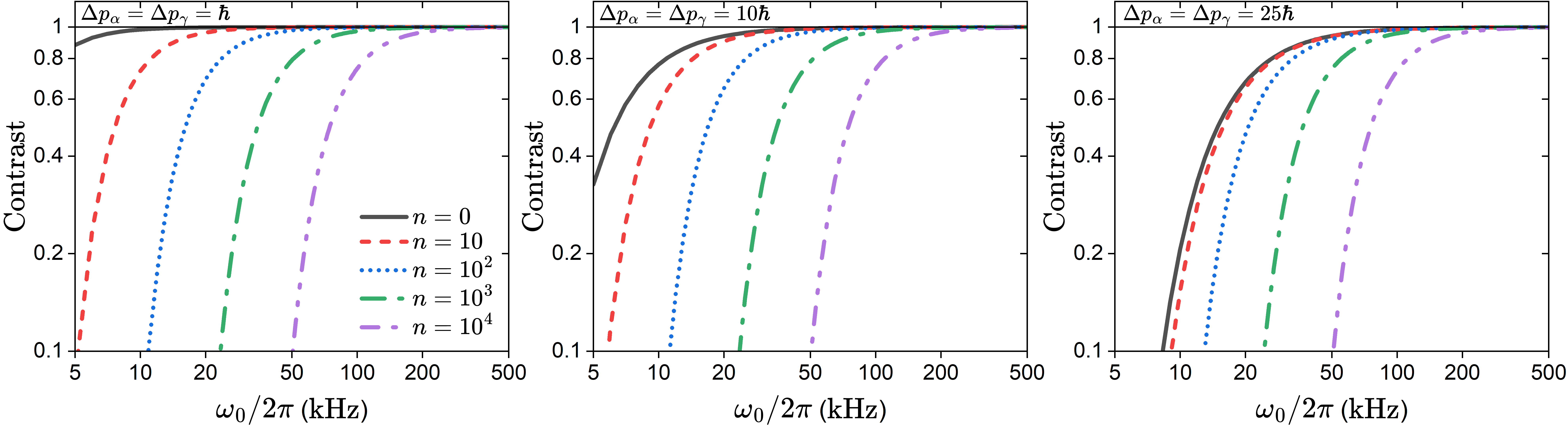}
    \caption{We show the spin contrast as a function of $\omega_0$ according to Eq. \eqref{eq:contrast_temprature} by varying $\Delta p_{(\alpha,\gamma)} = \{\hbar, 10\hbar, 25\hbar\}$ and for mass $m = 10^{-17} \ \text{kg}$ with an initial libration angle $\beta_0 = 0.01 \ \text{rad}$. We also  included the finite temperature case, $n = k_B T_{lib}/(\hbar \omega_0)$, from $T_{lib}=0 - 10^4$K. As seen in all three plots, with the increase in $\Delta p_{(\alpha,\gamma)}$, the occupation value for $n = 0 - 10^2$ exhibits a significant shift in contrast. This is indicated by the necessity of higher initial velocity to maintain high spin-contrast values. However, as the $n$ value increases, the need to shift to higher angular velocity becomes less significant, resulting in a stable trend for large values of $n$.  In all the examples, the spin is embedded at a location $d=10$ nm from the C.O.M, $L= 100$ nm, and the radius of the nanodiamond is $95.4$ nm.}
    \label{fig:contrast_temp}
\end{figure*}

\begin{figure*}
    \centering
    \includegraphics[width=1\linewidth]{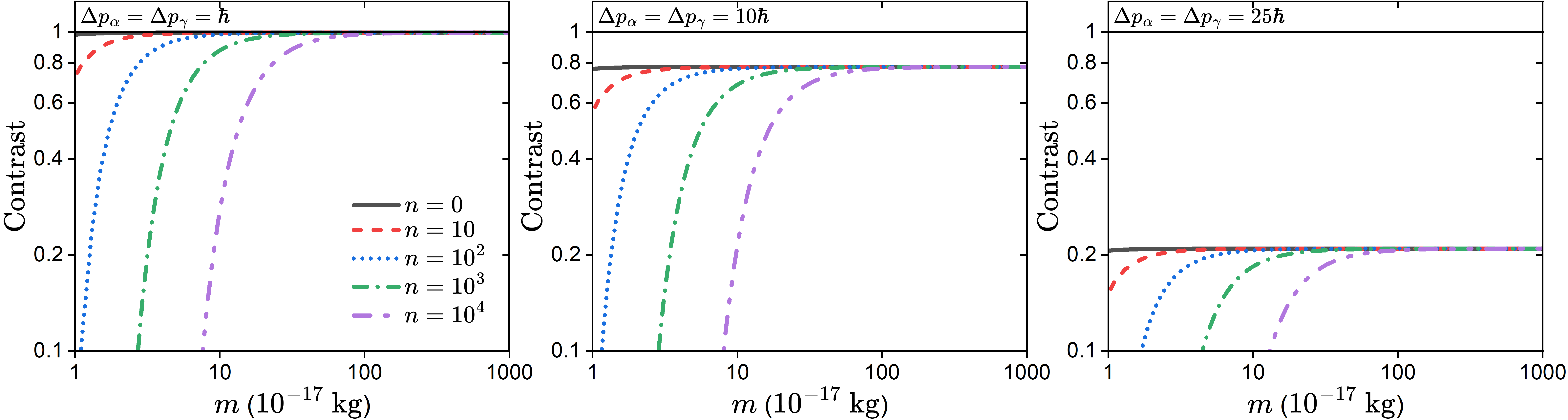}
    \caption{Same as Fig. \ref{fig:contrast_temp}, but right now we show the spin contrast as a function of mass ($m = 10^{-17} - 10^{-14} \ \text{kg}$), we take parameter $\omega_0 = 2\pi \times 10^4 \ \text{Hz}$. The variation of mass exhibits a different behaviour. In Figure \ref{fig:contrast_temp}, it is shown that for the three sets of parameters $\Delta p_{(\alpha,\gamma)}$, with a fixed mass $m = 10^{-17}$ kg and an initial velocity $\omega_0/(2\pi) = 10^4$ Hz, the spin contrast reaches approximately $C\sim 1, 0.8, 0.2$, respectively. These values represent the maximum spin contrast on this plot, showing the maximum contrast by variations in the occupation number/thermal libration temperatures. Good contrast is observed when the system is at a minimum mass and $\Delta p_{(\alpha,\gamma)}$, where an increase in $\Delta p_{(\alpha,\gamma)}$ causes a significant decrease in contrast values, reaching as low as 0.2 for $\Delta p_{(\alpha,\gamma)} = 25\hbar$.  Here  also, we consider the embedded spin is located at $d=10$ nm from the C.O.M. {We set the cylinder's height to $L=100$ nm, and its radius increases with mass, causing the shape to transition from a normal cylinder to a disk-shaped one. Nevertheless, since there is no significant difference between the two shapes (see Appendix \ref{appendix:long_cylinder}, where the disk-shaped configuration shows slightly better contrast), the overall contrast trend remains valid for cylinders in general, even when their shape changes.}}
    \label{fig:contrast_mass}
\end{figure*}

\begin{figure*}
    \centering
    \includegraphics[width=1\linewidth]{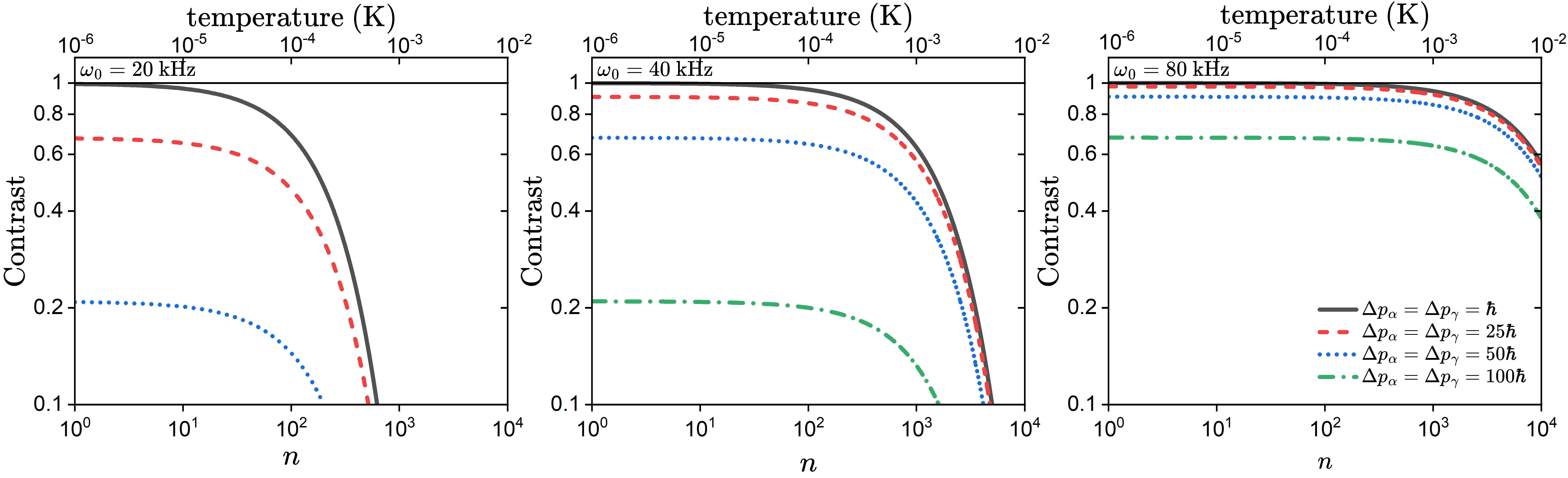}
    \caption{Another representation of spin contrast as a function of occupation number or libration temperature at fixed mass $m = 10^{-17}$ kg, with a variation of initial angular velocities  $\omega_0 = 20$, $40$ and $80$ kHz. In this plot, we aim to demonstrate the configuration values of $\Delta p_{(\alpha,\gamma)}$ and angular velocity that can be chosen to maintain a good spin contrast. We tested for $\Delta p_{(\alpha,\gamma)} = \hbar - 100\hbar$ and indicated that high $\Delta p_{(\alpha,\gamma)}$ values can be accepted if the initial angular velocity is also higher. It can be observed that an angular velocity of $\omega_0 = 80 \ \text{kHz}$ is required to enhance the spin contrast for configurations $\Delta p_{(\alpha,\gamma)} = 50 \hbar \ \text{and} \ 100\hbar$, and to reach a relatively high libration temperature ($T_{lib} \approx 10^{-2} \ \text{K}$). Note that this temperature is related to libration mode $\beta$, and should not be confused that of the C.O.M motional temperature.
    In all the examples, the embedded spin is located at $d=10$ nm from the C.O.M, $L= 100$ nm, and the radius of the nanorotor is $95.4$ nm.
    }
    \label{fig:contrast_n}
\end{figure*}

The quantum fluctuations lead to tiny deviation in the libration mode from its equilibrium position, while $p_\alpha^{\prime}$ and $p_\gamma^{\prime}$ are the same for the left and right arms of the interferometer, thus the overlap of the two coherent states $\abs{\prescript{}{\bar{\beta}_-}{\braket{\kappa_-}{\kappa_+}}_{\bar{\beta}_+}}$ is independent of the variable $p_\alpha^{\prime}$ and $p_\gamma^{\prime}$. Hence, we end up with a simple expression for the contrast.
\begin{small}
\begin{align}
    C & =\abs{\braket{\kappa_-}{\kappa_+}} \int \frac{d p_\alpha^{\prime} d p_\gamma^{\prime}}{2 \pi \Delta p_\alpha \Delta p_\gamma} e^{-\frac{\left(p_\alpha^{\prime}\right)^2}{2 \Delta p_\alpha^2}+\frac{i}{\hbar} \delta \alpha p_\alpha^{\prime}} e^{-\frac{\left(p_\gamma^{\prime}\right)^2}{2 \Delta p_\gamma^2}+\frac{i}{\hbar} \delta \gamma p_\gamma^{\prime}}  \nonumber \\
    & =\exp \left[-\frac{1}{2}\left(\delta \alpha^2 \frac{\Delta p_\alpha^2}{\hbar^2}+\delta \gamma^2 \frac{\Delta p_\gamma^2}{\hbar^2}\right)\right]\abs{\braket{\kappa_-}{\kappa_+}}
    \label{eq:contrast_not_complete}
\end{align}
\end{small}
We substitute Eqs.\eqref{eq:coherence_at_t_closed} and \eqref{eq:delta_X} into the inner product of the two coherence states in Eq. \eqref{eq:coherence_state_normalization}, we get the lower bound on the overlap between the two coherent states
\begin{align}
    \abs{\prescript{}{\bar{\beta}_{-}}{\braket{\kappa_-}{\kappa_+}}_{ \bar{\beta}_{+}}} &=\exp{-\frac{1}{2}\abs{\kappa_- - (\kappa_+ +\delta X)}^2}\\
    &> \exp \left[-\frac{1}{2}\left(\left|\kappa_-\right|+\left|\kappa_+\right|+\delta X\right)^2\right] \nonumber \\
    & >\exp \left[-\frac{1}{2} \frac{I \omega_0}{2 \hbar}\left(  \frac{I^2}{I_3^2} \frac{8 \mu B_0 \beta_0}{I \omega_0^2}\right)^2\right]
\end{align}
Therefore, the lower bound on the spin contrast due to the rotational degrees of freedom, Eq. \eqref{eq:contrast_not_complete}  can be given by:
\begin{small}
\begin{align}
    C > \exp\left(-\frac{\delta \alpha^2 \Delta p_\alpha^2}{2 \hbar^2}-\frac{\delta \gamma^2 \Delta p_\gamma^2}{2 \hbar^2}- \frac{I^4}{I_3^4} \frac{16 \mu^2 B_0^2 \beta_0^2}{\hbar I \omega_0^3}\right)
    \label{eq:contrast_non_T}
\end{align}
\end{small}
Next, we will compute the finite-temperature corrections in the libration mode, which affects the spin contrast in our case. We wish to see how deviation from the initial Gaussian state (zero temperature limit) will affect the spin contrast.

\section{Finite temperature and the libration mode}
\label{chapter:spin_contrast_with_temperature}

The spin contrast in Eq.\eqref{eq:contrast_non_T} is calculated in the zero-temperature limit, in which the initial state of the libration mode is assumed to be in the ground state of the harmonic trap. We can extend our contrast considering a finite temperature $T_{lib}$~\footnote{Note that this is the temperature, $T_{lib}$, of the libration d.o.f. This should not be confused with that of the phonon temperature, see~\cite{Xiang:2024zol}, or the motional temperature related to the C.O.M.~\cite{Deli2020,Piotrowski_2023}.}, in the initial state of the libration mode. The latter can be described by the density matrix operator in the coherent state basis, $\kappa$:
\begin{align*}
    \hat{\rho}_{th}=\int \frac{d^2 \kappa}{\pi} \frac{e^{-\frac{|\kappa|^2}{n}}}{n}\braket{\kappa}{\kappa}.
\end{align*}
This integral also includes the distribution function of the thermal state, with the occupation number in finite temperature given by
\begin{equation}
n \equiv \frac{k_B T_{lib} }{\hbar \omega_0}\,.
\end{equation}
 We define the final overlap between the two thermal coherent states as $\operatorname{tr}\left[\hat{D}\left(\zeta_-\right) \hat{\rho}_{th} \hat{D}^{\dagger}\left(\zeta_+\right)\right]$, where $\hat{D}(\zeta)$ is the displacement operator that satisfies $\hat{D}\left(\zeta_-\right)\ket{0}_{\bar{\beta}_0}=\ket{\kappa_-}_{\bar{\beta}}$ and $\hat{D}\left(\zeta_+\right)|0\rangle_{\bar{\beta}_0}=\ket{\kappa_+}_{\bar{\beta}}$. Consequently, according to \cite{steiner2024pentacene,Zhou:2024pdl}, we obtain the following expression:
\begin{align*}
    \operatorname{tr}\left[\hat{D}\left(\zeta_-\right) \hat{\rho}_{t h} \hat{D}^{\dagger}\left(\zeta_+\right)\right]=\exp \left[\varphi-|\Delta \zeta|^2\left(\frac{1}{2}+n\right)\right]
\end{align*}
The phase factor, $\varphi \equiv \left(\zeta_- \zeta_+^* - \zeta_-^* \zeta_+\right) / 2$, is an imaginary number, and thus does not affect the contrast. Consequently, when considering the thermal effects, $T_{lib}$ the equation should be reformulated, as follows~\cite{Zhou:2024pdl}
\begin{align}
    \left|\braket{\kappa_-}{\kappa_+}\right|_{t h} & =\exp \left[-\left(\frac{1}{2}+n\right)\left|\kappa_--\left(\kappa_+ +\delta X\right)\right|^2\right] \nonumber \\
    & >\exp \left[-\left(\frac{1}{2}+n\right)  \frac{I^4}{I_3^4} \frac{32 \mu^2 B_0^2 \beta_0^2}{\hbar I \omega_0^3}\right].
\end{align}
We can now define the lower bound on the spin contrast in the finite-temperature scenario as follows.
\begin{small}
\begin{align}
    C_{t h}>\exp \left[-\frac{\delta \alpha^2 \Delta p_\alpha^2}{2 \hbar^2}-\frac{\delta \gamma^2 \Delta p_\gamma^2}{2 \hbar^2}- \frac{I^4}{I_3^4}\frac{16(1+2 n) \mu^2 B_0^2 \beta_0^2}{\hbar I \omega_0^3}\right]. \label{eq:contrast_temprature}
\end{align}
\end{small}
The thermal noise causes a loss of contrast as a result of the overlap between the two libration states of the two arms of the interferometer. Since $\Delta p_\gamma = \Delta p_\alpha \cos{\beta_0}$, or $\Delta p_\gamma \approx \Delta p_\alpha$, for an extremely small angle of $\beta_0$, and from Eqs.\eqref{eq:double_dot_beta_minimal}, \eqref{eq:alpha_minimum}, and \eqref{eq:gamma_minimum}, we also have a mismatch in $\delta \beta \propto \omega_0^{-2}$ and $\delta \alpha = -\delta \gamma \propto \omega_0^{-1}$, which we will have to take into account in the above Eq. \eqref{eq:contrast_temprature}.

We can see the plots for the contrast as a function of $\omega_0$ for fixed mass $m=10^{-17}$ kg in Fig. \eqref{fig:contrast_temp} for different values of the uncertainties in $\Delta p_\alpha,~\Delta p_\gamma$. We see that contrast loss is prominent for higher libration temperatures (denoted here by the conversion $n=\hbar \omega_0/(K_BT_{lib})$). Still, we can compensate for the loss of contrast due to finite temperatures by appropriately increasing $\omega_0$. For an example, look at Fig. \eqref{fig:contrast_temp} the third panel; the contrast loss can be negligible if we ramp up to $\omega_0\sim 400 -500$ kHz. This is a remarkable lesson for nanorotors in the Stern-Gerlach interferometry.

For heavier masses, the contrast loss in spin coherence is evidenced in Figs. \ref{fig:contrast_mass}. If we combine this with Fig. \ref{fig:contrast_temp} with the fixed mass $m = 10^{-17}$ kg, at the point of $\omega_0/2\pi = 10 $ KHz, the contrast is approximately $C\sim 0.9, 0.8$, and $C_{th}\sim 0.2$, respectively, for cases $\Delta p_{\{\alpha,\gamma\}} =\hbar, 10\hbar$ and $25\hbar$. This value becomes the maximum contrast value and saturates for  $n=\hbar \omega_0/k_B T_{lib}$ and different masses, see Fig. \ref{fig:contrast_mass}. Showing that the stability of the nanodiamond is significantly influenced by both the initial angular velocity and the nanorotor's mass. An increasing mass of the nanorotor generally reduces spin contrast, primarily because of the increased inertia that dampens the libration motion. This effect is exacerbated at higher $\Delta p_{\{\alpha,\gamma\}}$ values. For another point of view, Fig. \ref{fig:contrast_n} shows the spin contrast as a function of occupation numbers and temperature of the libration mode.

Together, these figures highlight the interaction between mass, initial angular velocity, and thermal effects in determining the overall spin dynamics and stability of the nanorotor. Compared to the experimental parameters, for the case of the cylinder, the contrast is still close to $C\sim 1$ {for $\omega_0 \approx 40 \ \text{kHz}$, even for $n = 20$, which is compatible with the recent experimental cooling techniques of the libration mode; see \cite{Schafer2021}}.

\section{Conclusion}
\label{chapter:conclusion}

We briefly conclude what we discussed in this paper. First, we discuss the external and internal degrees of freedom of an axially symmetric nanorotor.  Although our analysis is very general, we fixed the parameters based on a nanodiamond with an NV centre defect embedded in it. For a generic nanorotor, we discussed how the strain of a crystal would affect the centre-of-mass motion in the presence of an external magnetic field, see Appendix \ref{appendix:E_term} We then examined the effect of rotation. Here we considered the dynamics of three Euler angles and how they evolve in the presence of the force imparted on a generic spin defect due to the external magnetic field.

We discussed the motion of the libration mode (see definition above) and studied the two spin eigenstates of spin $s=\pm 1$. We have already shown that by imparting rotation along the spin defect, we can exploit both spin states $|+1 \rangle $ and $|-1\rangle$~\cite{Zhou:2024pdl}. However, previous studies were performed in the context of a spherical system. In this paper, we study for the first time in the axially symmetric nanorotor. {We considered three possible configurations, the {long cylinder} (where the ratio of diameter-height is $D/L\approx0.1$), the normal cylinder (where the $D/L\approx1$) and the disk-shaped ($D/L\approx10$), we discuss the external and internal dynamics of those three shapes; see Fig.1.}

We then considered the one-loop matter wave interferometry of the cylinder by following the Stern-Geralch setup, where we followed the protocol of \cite{Marshman:2021wyk} of spin manipulation and the external magnetic field to create a spatial superposition for the mass range $m\sim 10^{-18}-10^{-15}$ kg. We considered the Humpty-dumpty problem for the spin coherence, which is inherently linked to the dynamical stability of the libration mode of the nanorotor. To obtain a large contrast, one has to ensure that we match the position and momentum of the nanorotators after the completion of the interference. Moreover, we must also ensure that all three Euler angles match. 

To obtain maximum rotational stability, the initial angular velocity $\omega_0$ should point to the principal axis with the largest inertia moment; see Fig.(\ref{fig:beta_equil(t)}). A higher value of $\omega_0$ can reduce the amplitude of the spin-axis libration and get a better spin contrast of the SGI, see Fig.(\ref{fig:contrast_temp}). Remarkably, for the magnetic field configuration we applied in this paper, the value of $\omega_0$ should be limited in the range from kHz to MHz, to avoid possible spin flips (Majorana and Rabbi oscillations due to the degeneracy between the states $|+1\rangle $ and $\ket{-1}$), the latter is an effect due to the Einstein-de Haas effect. The Einstein-de Haas effect arises primarily from the $\mathcal{D}$-term constraint at the spin-1 defect. Such an effect will be impossible to get in spin-1/2 systems, as the zero-point energy splitting in such a case yields the $\mathcal{D}$-term zero, e.g. $\mathcal{D}=0$.

Furthermore, we also considered the effect of the thermal state on the degree of freedom of liberation. We studied cases for different $\omega_0$ to alter the initial temperature of the motion cooling of the libration mode. We found that for reasonable $\omega_0\sim 80$ kHz we can tolerate temperatures up to $10^{-4}$K without sacrificing the spin contrast, which means that we do not require the libration mode to cool all the way in an experiment. 

The future outlook is very interesting; first of all, our protocol provides a way to control the nanorotor's rotation in the presence of the external magnetic field. It provides gyroscopic stability like a rotating bullet as it carves its trajectories for the left and right parts of the spatial interferometer.
{Second, in the cylindrical nanorotor, our protocol provides gyroscopic stability for the normal and disk-shaped cylinder case; see Fig.~\ref{fig:1} (b), better than the {long cylinder} case for same masses, the stability is getting worse at lower mass ($\sim10^{-18}-10^{-17}$ kg). Also in the normal and disk-shape regime, as we increase the mass, the mismatch in the libration mode significantly decreases. As we increase the mass of the nanorotor, the gyroscopic stability becomes better and better, as we can see from the mismatch of the angular modes, see Fig.\ref{fig:alpha_gamma}(a) and (b). 
Finally, we see that the spin contrast due to the libration mode can be completely ameliorated in the zero- and finite-temperature cases, which is a great outlook for the QGEM experiment and subsequent experiments with lower masses and smaller superposition sizes. As we conclude this study, we came across an intriguing article that explores different aspects of levitating rigid bodies, see~\cite{Wachter2025sbn}.}

Complexifying our scenario will involve considering multiple spin defects and performing a similar analysis to show that gyroscopic stability can maintain its advantages. This study is slightly out of the context of this paper, but future investigation is a must to realise the QGEM experiment in a lab.

\begin{acknowledgments}
R.R. is supported by Beasiswa Indonesia Bangkit, the Ministry of Religious Affairs of the Republic of Indonesia (Kemenag), and the Indonesia Endowment Fund for Education (LPDP) of the Ministry of Finance of the Republic of Indonesia. T.Z. is supported by the China Scholarship Council (CSC). SB
thanks EPSRC grants EP/R029075/1, EP/X009467/1,
and ST/W006227/1. S.B. and A.M.'s research is funded by the Gordon and Betty Moore Foundation through Grant GBMF12328, DOI 10.37807/GBMF12328. This material is based on work supported by the Alfred P. Sloan Foundation under Grant No. G-2023-21130

\end{acknowledgments}

\bibliography{apssamp}

\appendix
\section{Feschbach Formalism}
\label{appendix:feshbach}
Recall the spin Hamiltonian Eq.~\eqref{eq:matrix_form_hamiltonian_with_spin}, 
\begin{align*}
{H}_s
    & = \begin{pmatrix}
        \mu B_\parallel & \mu \frac{B_\perp}{\sqrt{2}}e^{-i\gamma} & E \\
        \mu \frac{B_\perp}{\sqrt{2}}e^{i\gamma} & -\mathcal{D} & \mu \frac{B_\perp}{\sqrt{2}}e^{-i\gamma} \\
        E & \mu \frac{B_\perp}{\sqrt{2}}e^{i\gamma} & -\mu B_\parallel
    \end{pmatrix}
    + \frac{\mathcal{D}}{3} \mathbb{I},
\end{align*}
We apply the projection operators 
\begin{eqnarray}
P = \ket{1}\bra{1} + \ket{-1}\bra{-1}, \
Q = \mathbb{I} - P = \ket{0} \bra{0},
\label{eq:P_and_Q}
\end{eqnarray}
to separate between the state $\ket{\pm 1}$ and the level $\ket{0}$. With the help of projection operators, we will derive an effective $2 \times 2$ Hamiltonian for the levels $\ket{\pm 1}$ states. Especially when the energies and rates of changes involved in the interaction with the external magnetic field are much smaller than the zero-field splitting $\mathcal{D}$, we can concentrate on the reduced Hamiltonian.

 To determine the system's eigenstates and eigenvalues, the Schrodinger equation $i\hbar \partial_t \psi= H \psi$, can then be separated with the help of projection operators; 
 $$\mathcal{E}P\psi = PHP \cdot P\psi + PH_sQ \cdot Q\psi $$ and $$\mathcal{E}Q\psi = QH_sQ \cdot Q\psi + QH_sP \cdot P\psi$$, and we can obtain the relation
\begin{align}
    \mathcal{E}P\psi & = \left( PH_sP + PH_sQ  \frac{1}{\mathcal{E}-QH_sQ} QH_sP \right) P\psi.
\end{align}
We can reasonably approximate when $QH_sQ$ has a larger magnitude compared to the other parts of the Hamiltonian. Hence, for energies $|\mathcal{E}| \ll |QH_sQ|$, we can approximate the energy eigenstates and eigenvalues by taking $\mathcal{E} \rightarrow 0$~\footnote{For the Hamiltonian in Eq. \eqref{eq:matrix_form_hamiltonian_with_spin} the expression for every projection operators are $PH_sP = \mu B_\parallel\ket{1}\bra{1}+E\ket{-1}\bra{1}+E\ket{1}\bra{-1}-\mu B_\parallel E\ket{-1}\bra{-1}$, $PH_sQ = (QH_sP)^\dagger = \mu B_\perp/\sqrt{2} (e^{-i\gamma} \bra{1}\ket{0}+e^{i\gamma}\ket{-1}\bra{0})$, and $QH_sQ = \mathcal{D} \ket{0}\bra{0}$. In our case, $\mathcal{D} \gg \mu B, ~E$, making the component $\abs{QH_sQ}$ significantly larger than the other projection operators}. So we have an effective Hamiltonian given by\cite{Band2022}. 
\begin{align}
    H_{eff} = PH_sP - PH_sQ  \frac{1}{QH_sQ} QH_sP.
    \label{eq:H_effective_transform}
\end{align}
Now, we can substitute the spin Hamiltonian, Eq. \eqref{eq:matrix_form_hamiltonian_with_spin}, and the definition of $P$ and $Q$ in Eq.~\eqref{eq:P_and_Q}, we have
\begin{align}
    PH_sP = & \ \mu B_\parallel \ket{1}\bra{1} + E \ket{-1}\bra{1} \nonumber \\
    & + E \ket{1}\bra{-1} - \mu B_\parallel \ket{-1}\bra{-1} \\
    PH_sQ = & \ \frac{\mu}{\sqrt{2}} (B_\perp e^{-i\gamma} \ket{1}\bra{0} + B_\perp e^{i\gamma} \ket{-1}\bra{0}) \\
    QH_sQ = & \mathcal{D} \ket{0} \bra{0}
\end{align}
so, the effective Hamiltonian becomes:
\begin{align}
    H_{eff}  =\ & \mu B_\parallel \ket{1}\bra{1} + E \ket{-1}\bra{1} + E \ket{1}\bra{-1} - \mu B_\parallel \ket{-1}\bra{-1} \nonumber \\
    & + \frac{\mu^2}{2\mathcal{D}} (B^2_\perp \ket{1}\bra{1} + B^2_\perp \ket{-1}\bra{-1}) \nonumber \\
    & + B^2_\perp e^{-2i\gamma} \ket{1}\bra{-1} + B^2_\perp e^{2i\gamma} \ket{-1}\bra{1} + \frac{\mathcal{D}}{3} I, \nonumber
\end{align}
or in the matrix form
\begin{align}
    H_{eff} = 
    \begin{pmatrix}
        \mu B_\parallel & \epsilon(B_\perp)^* \\
        \epsilon(B_\perp) & -\mu B_\parallel
    \end{pmatrix}
    + \frac{1}{3} \left( \mathcal{D} + \frac{3\mu^2 B_\perp^2}{2\mathcal{D}} \right) I
\end{align}
where $\epsilon(B_\perp) = E + \frac{\mu^2}{2D} B_\perp^2 e^{2i\gamma} $. 
Now, we can easily calculate the eigenvalue of $H_{eff}$ which leads to the force that contributed to the spin.

\section{Evolution of Time of The Rotation}
\label{appendix:A}
 The time evolution of an operator $A$ in the Heisenberg picture can be written as
\begin{align}
    \dot{A} = \frac{i}{\hbar} [H,A]\, ,
\end{align}
as long as A does not have an explicit time dependency. This is applicable here since the operators for angular momentum, spin, and principal axes are not explicitly time-dependent. The spin is embedded in a nanorotor. The principal axes can be considered operators because they are essentially unit vector position operators aligned with the ND. Referring to the Hamiltonian given in eq.\eqref{eq:hamiltonian_spin_only}, we obtain 
\begin{align}
    \dot{\mathbf{S}} & = \frac{i}{\hbar} [H_s,\mathbf{{S}}], \\
    \dot{\mathbf{L}} & = \frac{i}{\hbar} [H_s,\mathbf{L}], \\
    \dot{\hat{n}}_i & = \frac{i}{\hbar} [H_s,\hat{n}_i] \, .
\end{align}

The non-zero commutation relation for every term of $H_s$ to $S$, $L$, and $\hat{n}_i$, are given by:
\begin{align}
    \dot{\mathbf{S}} & = \frac{\mu}{\hbar} [{\mathbf{S}} \cdot \mathbf{B},\mathbf{{S}}] + \frac{i}{\hbar} [H_{ZFS},\mathbf{{S}}], \label{eq:evol_S}\\
    \dot{\mathbf{L}} & = \frac{i}{\hbar} [H_{ZFS},\mathbf{L}], \label{eq:evol_L} \\
    \dot{\hat{n}}_i & = \frac{i}{2\hbar} \sum_k \left[\frac{L^2_k}{I_k},\hat{{n}}_i\right]. \label{eq:evol_n}
\end{align}
where $i = {1,2,3}$ denotes the index for the rotating frame. As observed, the gravitational term in the Hamiltonian is null because $\mathbf{r}$ is zero in the co-moving frame. Intuitively, this can be understood as gravity affecting the center of mass of the nanorotor and hence exerting no torque owing to Earth's gravity. This explanation holds if the ND is levitated, as in the scenario of diamagnetic levitation, see~\cite{Elahi:2024dbb}. We are aware of the commutators: $[S_i , S_j]  = i \epsilon_{ijk}S_k, \ [L_i, L_j]  = i\hbar \epsilon_{ijk} L_k, \ [\hat{e}_i, L_j]  = i\hbar \epsilon_{ijk} e_k,$
where $\hat{e}_i$ components are relative to the center of the nanorotor, and $\epsilon_{ijk}$ is the Levi Civita symbol.

The spin operator $S_j = \mathbf{S}\cdot \hat{{n}}_j$ is given in the rotating frame. For eq.\eqref{eq:evol_S}, $[{\mathbf{S}} \cdot \mathbf{B},\mathbf{S}]  = [S_i B_i,\mathbf{S}] + [S_j B_j,\mathbf{S}] + [S_k B_k,\mathbf{S}] = -i \mathbf{B} \cross \mathbf{S}$.

For Eq.\eqref{eq:evol_n}, we used relations $[A^2,B] = A[A,B]+[A,B]A$, and $\{A,B\} = AB+BA = 2AB - [A,B]$, so
\begin{align}
    \frac{i}{2\hbar} \sum_k \left[\frac{L^2_k}{I_k},\hat{{n}}_i\right]  =  \boldsymbol{\omega}\cross\hat{\hat{n}}_i+i \Gamma \hat{{n}}_i,
\end{align}
where $\omega_k = L_k/I_k$, and $\Gamma = \sum_k {\hbar}/{2I_k}$.

For equations \eqref{eq:evol_S} and \eqref{eq:evol_L}, we can use the relation; $[\mathbf{S}\cdot \hat{{n}}_j, \mathbf{S}] = -i \hat{{n}}_j \cross \mathbf{S}$ and $[\mathbf{S}\cdot \hat{{n}}_j, \mathbf{L}] = i \hbar \hat{{n}}_j \cross \mathbf{S},$
\begin{figure}
    \includegraphics[width=0.85\linewidth]{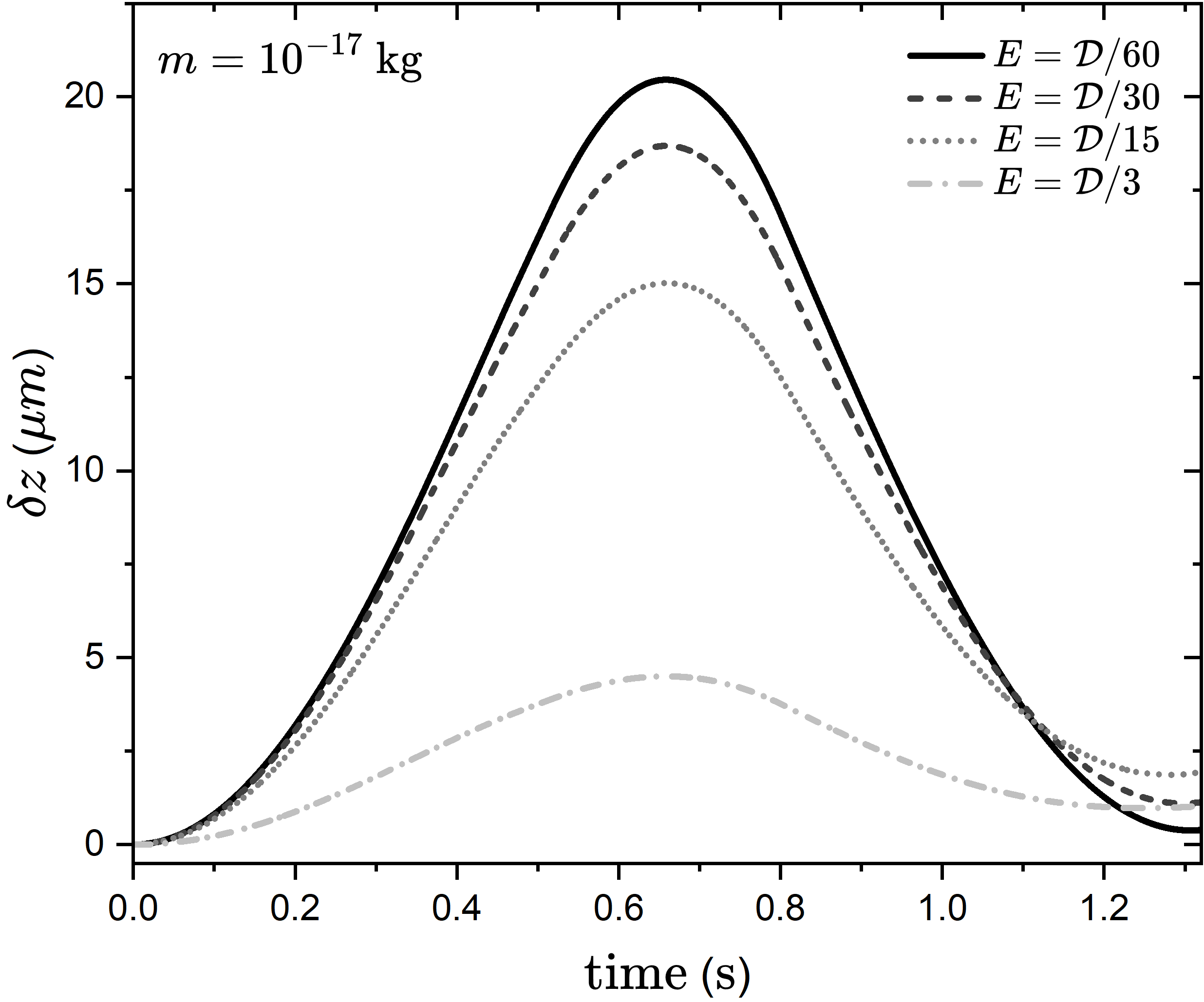}      
    \caption{The size of the superposition with respect to $E$ values is determined using the solution from the equation of motion, Eq. \eqref{eq:z_eom}. Here, $\mathcal{D}=2.87 h \times 10^9$ Hz (similar to that of a nanodiamond), with a mass of $m=10^{-17} \text{kg}$. The magnetic field profile from Eq. \eqref{eq:magnetic_field_profile} is applied, and $\chi_\rho$ is set to $-6.2\times 10^{-9} \text{m}^3/ \ \text{kg}$. Introducing the parameter $E$ into the Hamiltonian ZSF affect the superposition size. The maximum size is around $20 \ \mu m$ when $E=\mathcal{D}/60$, comparable to the scenario when $E=0$, and reduces to approximately $5 \ \mu m$ at $E=\mathcal{D}/3$, which is the mathematical upper bound for $E$. In typical nanodiamonds, the $E$ term is much smaller than the $\mathcal{D}$ term, with an experimental value of approximately 10 MHz \cite{Hoang2016}. For our case, we assume an initial angular velocity of $\omega_0 = 2\pi\times 10 \ \text{GHz}$, and the embedded spin is located at $d=10 \ \text{nm}$ from the CM with angle with respect to $\hat{n}_3$ being $\alpha' = \pi/6$, and the cylinder length and radius are $L= 100$ and $95.4$ nm, respectively.}
    \label{fig:E_term}
\end{figure}
\begin{figure*}
    \centering
    \includegraphics[width=1\linewidth]{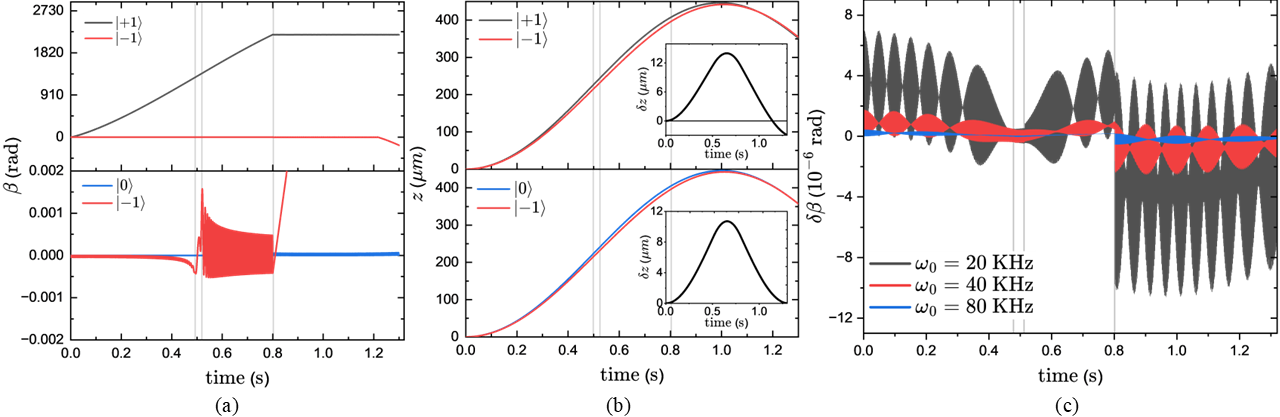}
    \caption{The solution of equation (a) \eqref{eq:beta_no_initial_velocity} and (b) \eqref{eq:z_no_initial_velocity}, with $\omega_0 = 0$ for two pair NV-spin state cases $s$ at \{+1,-1\} and \{0,-1\}, with +1 represent as black line -1 for red line and 0 for blue line, and superposition size $\delta z$ for green line. In this case, The magnetic field profile that we used for each arm of the interferometer as a function of time has constant parameters $B_0$, $B_1$, and $\eta$, set at values of $10^{-2} T$, $10^{-4} T$, and $0.45 \times 10^2 T/m$ respectively, for the time conditions $\tau_1 = 0.482 \ \text{s}$ and $\tau_2 = 0.514 \ \text{s}$, as referred to in Ref. \cite{Marshman2022}. As we can see in (a), the behaviour is suitable as predicted by \cite{japha2022role, Japha:2022phw}. Without $\omega_0$, the nanorotor with the chosen state $s = \{0,-1\}$ is more stable than $\{+1,-1\}$. In the range time $\tau_2<t\leq t_{flip}$, the state $\ket{-1}$ at case $s=\{0,-1\}$ experience libration frequency as $\omega=\sqrt{\mu B_z/I}$, and for $t>t_{flip}$ the spin flip to $\ket{0}$ state and freely rotate. In (b), the spatial trajectory is shown, and there is no significant pattern difference here. The maximum superposition size is about 11 $\mu m$ for $s= \{0,-1\}$ and about 13 $\mu m$ for $s= \{+1,-1\}$. However, due to the large value of $\beta$ in the case of $s= \{+1,-1\}$, the trajectories of the two arms overlap after $t=1.15$ s. We introduce $\omega_0$ on (c), shows The mismatch of libration angle $\delta\beta$ from the solution to equation \eqref{eq:double_dot_beta_solve} for spin states $\ket{\pm1}$ with varying initial angular velocities from $\omega_0 = 20 - 80 \ \text{kHz}$. This angular mismatch $\delta\beta$ shows that increasing the initial angular velocity reduces the amplitude of the libration motion, indicating that higher velocities make the rotation of the nanorotor more stable. In this case, we used {normal cylinder shape} configuration with the mass of $m = 10^{-17} \ \text{kg}$ and the spin defect is located at $d = 10$ nm with angle $\alpha' = \pi/6$ and  cylindrical length and radius are $L = 100$ nm and $95.4$ nm, respectively.}
    \label{fig:zero_and_-1}
\end{figure*}
\begin{figure*}
    \centering
    \includegraphics[width=0.9\linewidth]{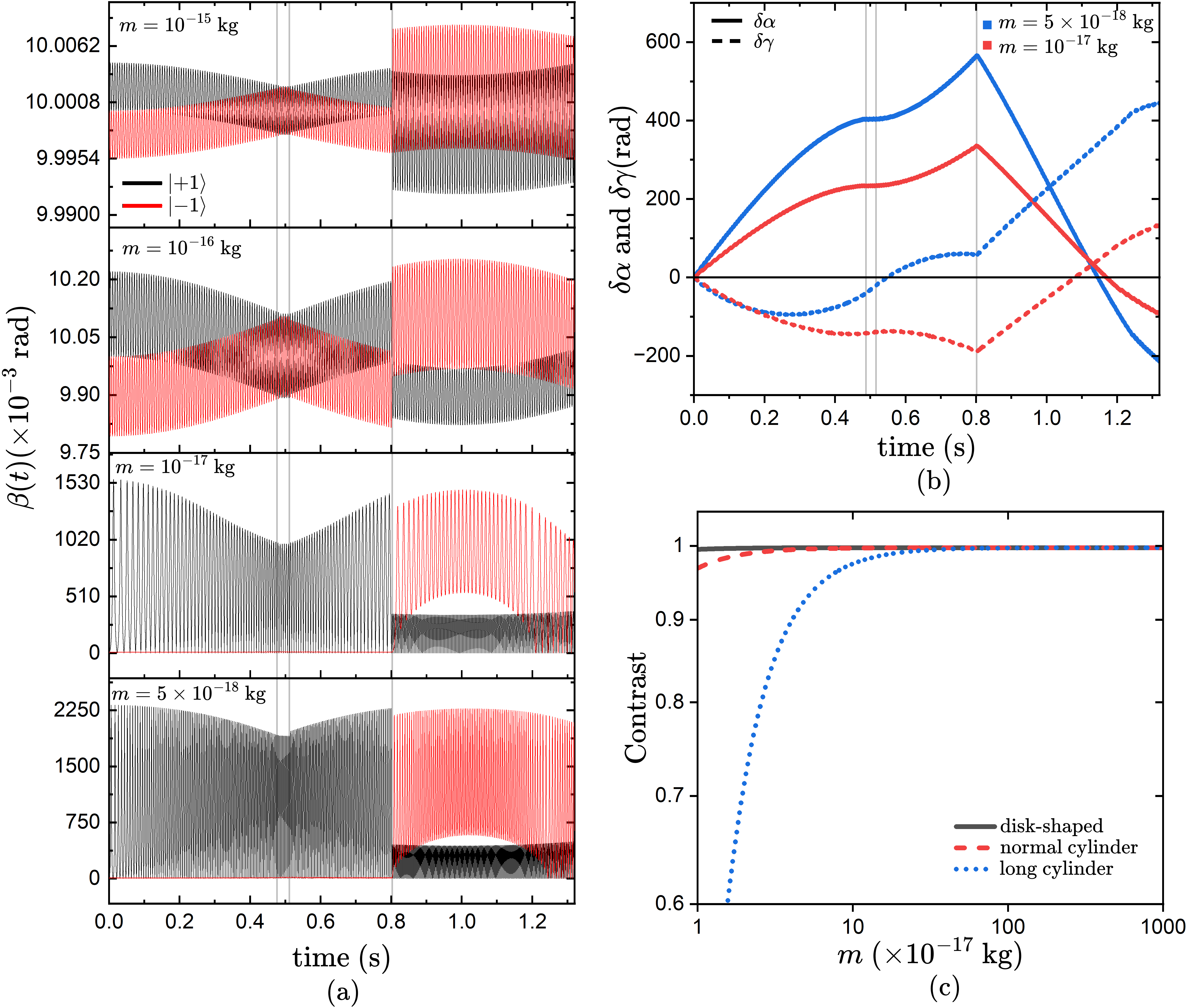}
    \caption{{(a) The solution to Eq. \eqref{eq:double_dot_beta_solve} for long cylinder case $D/L =0.1$ using similar parameter from Fig. (\ref{fig:beta(t)}), (b) solutions of \eqref{eq:dot_alpha_solve} and \eqref{eq:dot_gamma_solve} for the long cylinder case using a similar parameter from Fig (\ref{fig:alpha_gamma}), and (c) the contrast comparison for three possible cylinder shapes using the configuration $\Delta p_{(\alpha,\gamma)}=\hbar$, occupation number $n=0$ and initial rotation $\omega_0=2\pi\times10^4 \text{ Hz}$. It appears that the shape of the cylinder significantly influences the libration mode $\beta$ in small masses ($\sim10^{-18} - 10^{-17}$ kg) when the shape is long cylinder, where the libration angle amplitude becomes very large and uneven in both spin states. In contrast, in the same shape, the amplitude stabilises during the transition to larger masses ($10^{-16} - 10^{-15}$ kg). Consequently, for $\delta\alpha$ and $\delta\gamma$, in low masses ($5\times 10^{-18}$ and $10^{-17}$ kg), when the cylinder is long-shaped, $\delta\alpha$ and $\delta\gamma$ become asymmetric. The relationship $\delta\alpha \approx-\delta\gamma$ becomes a very important feature in determining the contrast of SGI. Considering contrast conditions, objects with the same mass but different shapes can exhibit different contrasts. For low mass, a normal cylinder and a disc shape are much better compared to a long cylinder. 
    }}
    \label{fig:long_cylinder}
\end{figure*}
and for $[H_s,\mathbf{S}]$ and $[H_s,\mathbf{L}]$ are given by:
\begin{align}
    [H_{ZFS},\mathbf{S}] = &  \left[\mathcal{D} \left( S_k^2-\frac{S(S+1)}{3}  \right) +  E (S_i^2 - S_j^2),\mathbf{S} \right], \nonumber \\
    = &  -i\mathcal{D}(S_k (\hat{{n}}_k \cross \mathbf{S}) +(\hat{{n}}_k \cross \mathbf{S}) S_k) \nonumber\\
    & -iE(S_i(\hat{{n}}_i \cross \mathbf{S}) + (\hat{{n}}_i \cross \mathbf{S}) S_i) \nonumber \\
    & +iE(S_j (\hat{{n}}_j \cross \mathbf{S})+(\hat{{n}}_j \cross \mathbf{S})S_j), \nonumber \\
    [H_{ZFS},\mathbf{L}] = & \left[\mathcal{D} \left( S_k^2-\frac{S(S+1)}{3}  \right) +  E (S_i^2 - S_j^2),\mathbf{L} \right] \nonumber \\
    = & \ i\hbar \mathcal{D}(S_k (\hat{{n}}_k \cross \mathbf{S}) + (\hat{{n}}_k \cross \mathbf{S}) S_k) \nonumber \\
    &+ i \hbar E(S_i (\hat{{n}}_i \cross \mathbf{S}) +(\hat{{n}}_i \cross \mathbf{S}) S_i)\nonumber \\
    &-i \hbar E(S_j (\hat{{n}}_j \cross \mathbf{S})+(\hat{{n}}_j \cross \mathbf{S})S_j), 
\end{align}
we can simplify with case $E=0$ and NV-spin align in $\hat{n}_3$ therefore $S_\parallel = \mathbf{S}\cdot\hat{n}_3$ so that
\begin{align}
    [H_{ZFS},\mathbf{S}] = &  -i\mathcal{D}(S_k (\hat{{n}}_k \cross \mathbf{S}) +(\hat{{n}}_k \cross \mathbf{S}) S_k) \nonumber\\
    = & -i [\mathcal{D}S_\parallel^2, \mathbf{S}], \\
    [H_{ZFS},\mathbf{L}] = & \ \hbar \mathcal{D}(S_k (\hat{{n}}_k \cross \mathbf{S}) + (\hat{{n}}_k \cross \mathbf{S}) S_k) \nonumber \\
    = & \ i\hbar [\mathcal{D}S_\parallel^2, \mathbf{S}] .
\end{align}
Hence, we have the following relation $[H_{ZFS},\mathbf{L}] = -\hbar [H_{ZFS},\mathbf{S}].$ Now, we can write down the complete solutions for the Heisenberg picture are then given by:
\begin{align}
    \dot{\mathbf{S}} & = \frac{\mu}{\hbar} (\mathbf{B} \cross \mathbf{S}) + \frac{i}{\hbar} [H_{ZFS},\mathbf{{S}}], 
    \label{eq:evol_sol_S} \\
    \dot{\mathbf{L}} & = -i [H_{ZFS},\mathbf{S}], \label{eq:evol_sol_L} \\
    \dot{\hat{n}}_i & = \boldsymbol{\omega}\cross\hat{{n}}_i+i \Gamma \hat{{n}}_i. \label{eq:evol_sol_n}
\end{align}
It is important to note that the magnetic susceptibility is not mentioned here because it influences only the spatial movement, not the rotations of the diamond, meaning it does not apply any external torque. Nevertheless, these terms must be included in the original Hamiltonian to fully account for the operation of the interferometer, as they impact the trajectory of the objects.

Since the principal axes form a basis that co-rotates with the ND, it is useful to define the spin, angular momentum, and the magnetic field in the co-moving frame: $S_i = \mathbf{S}\cdot\hat{{n}}_i, \ L_i = \mathbf{L}\cdot\hat{{n}}_i, \ B_i = \mathbf{B}\cdot\hat{{n}}_i$.
The time evolution of the spin components in this frame can be written as $\dot{S}_i =   \dot{\mathbf{S}} \cdot \hat{{n}}_i + \mathbf{S} \cdot \dot{\hat{{n}}}_i$, or
\begin{align}
    \dot{S}_i = \frac{i}{\hbar} [\mathcal{D}S_\parallel^2,S_i]+ \epsilon_{ijk} \left(\frac{\mu}{\hbar} B_j S_k +  S_j \omega_k \right) + i \Gamma s_i .
    \label{eq:Si_time_derivative_A}S
\end{align}
The assumption that the spin remains aligned with the crystal now manifests itself as an additional constraint $\dot{S}_i = 0$. Rewriting, $[\mathcal{D}S_\parallel^2, S_i] = i\epsilon_{ijk} \left( \mu B_j S_k + \hbar S_j \omega_k \right)$, and similar approach for the angular momentum, with angular velocity $\omega_k = L_k/I_k $ yields $\dot{L}_i =  \dot{\mathbf{L}} \cdot \hat{{n}}_i + \mathbf{L} \cdot \dot{\hat{{n}}}_i$, or
\begin{align}
     \dot{L}_i = \epsilon_{ijk} \left(\mu B_j S_k + {\hbar} S_j \frac{L_k}{I_k}+L_j\frac{L_k}{I_k} \right).
     \label{eq:Li_time_derivative_A}
\end{align}
The rotational evolution of the angular momentum of the nanorotor can be observed, encompassing the effects of Zeeman torque and the Einstein-de Haas effect.

\section{E term and Zero-field Splitting}
\label{appendix:E_term}
We present the full expression of the differential equation for the z-trajectory as
\begin{align}
    \Ddot{z} & = \mp  \frac{\mu^2 B_\parallel}{m \sqrt{(\mu B_\parallel)^2+\abs{E}^2}} \frac{\partial B_\parallel}{\partial z} - \frac{\chi_\rho}{2\mu_0} \frac{\partial \mathbf{B}^2}{\partial z}, 
\end{align}
which can be solved numerically with initial conditions $z(0) = \dot{z}(0) = 0$. Together with the rotating equation of motion, we include equation Eq.~\eqref{eq:double_dot_beta_solve}, resulting in Fig.\ref{fig:E_term}.

The presence of the parameter $E$ in the Hamiltonian ZSF contributes to a notable decrease in the extent of the superposition. In actual nanodiamonds, the term is usually much smaller than the $\mathcal{D}$ term (with experimental values around $E \sim 10$ MHz \cite{Hoang2016}), yet, despite its small magnitude, it still impacts the coherence and behaviour of quantum states within the nanodiamond. To comprehend the subtle effects on the system's dynamics and performance, this influence needs to be incorporated into precise modelling and experimental assessments.

\section{With and without initial rotation}
\label{appendix:B}

In this section, we examine the evolution of the libration angle $\beta(t)$. We alter the initial velocity parameters and take $m = 10^{-17} \ \text{kg}$ and $\omega_0$ that vary from $0$ to $80 \ \text{GHz}$. Building on the insights from \cite{Zhou:2024pdl,Japha:2022phw} concerning the stability of spin states $\ket{-1}$ and $\ket{0}$ when $\omega_0 = 0$, we solve Eq. \eqref{eq:beta_no_initial_velocity}. Furthermore, we will explore and compare these findings with $\ket{+1}$ and $\ket{-1}$ for the $\omega_0\neq 0$ values by solving Eq.~\eqref{eq:double_dot_beta_solve}, as shown in Fig. \ref{fig:zero_and_-1}.

In Fig.~\ref{fig:zero_and_-1}, we examine the scenario in which the angular frequency \(\omega_0\) is zero. In this situation, the mass of the system is specified as $m = 10^{-17}$ \text{kg}, and the characteristic length and radius are $100$ and $95.4$ \text{nm}. Throughout the period $\tau_2 < t \leq t_{flip}$, the system assumes the state $\ket{-1}$, showing a libration frequency of $\omega = \sqrt{\mu B_z / I}$. This libration is a result of interactions determined by the outlined parameters. However, for $t > t_{flip}$, there is a crucial transition in which the spin transitions to the $\ket{0}$ state, enabling the nanorotor to rotate freely, free from the constraints of the former state. Let us compare it with the spin states $s = \{+1,-1\}$. The spin $s=+1$ state appears to be very unstable with a very large rotation $\beta$, resulting in the
mismatch in the libration mode when the interferometer is closed
at $t = 1.15$ s. The stable spin states are $s=\{0, -1\}$ and are consistent with the predictions made by \cite{japha2022role, Japha:2022phw}.

Furthermore, the analysis shown in Fig.~ \ref{fig:zero_and_-1}(c) highlights a key element of the system's dynamics. The angular mismatch \(\delta\beta\) shows that an increase in the initial angular velocity leads to a reduction in the amplitude of the libration motion. This finding indicates that ramping up initial angular velocities enhances the rotational stability of the nanorotor. Consequently, controlling the initial velocity is essential for achieving a more stable rotation in the internal states of the nanorotor. 

\section{Rotation of long cylinder}
\label{appendix:long_cylinder}
In the case of nanodiamonds shaped as cylinders, there are three possibilities: long cylinders, normal cylinders, and short disk-shaped cylinders. Specifically, elongated cylinders, in which the ratio $D/L \approx 0.1$, have a much higher libration amplitude compared to other shapes. They also tend to be unstable if the mass is small, around $m \approx 10^{-18} - 10^{-17}$ kg, affecting their contrast value. A clearer illustration can be seen in Fig. \ref{fig:long_cylinder}.

\end{document}